\documentclass[%reprint,
prl,
superscriptaddress, twocolumn,
 amsmath,amssymb,
 aps,
]{revtex4-2}

\usepackage{graphicx}% Include figure files
\usepackage{color}
\usepackage{subfigure}
\usepackage{float}
\usepackage{amsmath}
\usepackage{ulem}
\usepackage{dcolumn}% Align table columns on decimal point
\usepackage{bm}% bold math
\usepackage{hyperref}% add hypertext capabilities
\begin{document}

% \preprint{APS/123-QED}
\newcommand{\thistitle}{
Quantum Melting of Spin-1 Dimer Solid Induced by Inter-chain Couplings
%Quantum Melting of Spin-1 Dimerized Order Induced by Inter-chain Coupling
}

\title{\thistitle}
% Optional Title: (1) Melting of Dimerized Order in Weakly Coupled Chains Limits (2) Easy-Melting Nature (Instability) of Dimerized Order in Spin-1 Bilinear-Biquadratic Heisenberg ModeL (3) Quantum Melting Transition of Dimer Order Induced by Weak Inter-chain Couplings in $S=1$ Bilinear-Biquadratic Heisenberg Model

% Force line breaks with \\
%\thanks{A footnote to the article title}%
%Quantum Melting Transition of Dimer Order Induced by Weak Inter-chain Couplings in $S=1$ Bilinear-Biquadratic Heisenberg Model
\author{Yi Xu}
 % \altaffiliation[Also at ]{Department of Physics and Astronomy, Rice University}%Lines break automatically or can be forced with \\
 \email{yx51@rice.edu}
\affiliation{Department of Physics and Astronomy, Rice University, Houston, TX 77005, USA}

\author{Tianfu Fu} 
%\email{tf13@rice.edu}
\affiliation{Department of Physics and Astronomy, Rice University, Houston, TX 77005, USA}
 
\author{Juraj Hasik}
 \email{j.hasik@uva.nl}
 \affiliation{Laboratoire de Physique Théorique UMR5152, C.N.R.S. and Université de Toulouse, 118 rte de Narbonne, 31062 Toulouse, FRANCE}
 \affiliation{Institute for Theoretical Physics and Delta Institute for Theoretical Physics, University of Amsterdam, Science Park 904, 1098 XH Amsterdam, The Netherlands}

\author{Andriy H. Nevidomskyy}
 \email{nevidomskyy@rice.edu}
\affiliation{Department of Physics and Astronomy, Rice University, Houston, TX 77005, USA}

\date{\today}

\begin{abstract}
Dimerized valence bond solids  appear naturally in spin-1/2 systems on bipartite lattices, with the geometric frustrations  playing a key role both in their stability and the eventual `melting' due to quantum fluctuations. Here, we ask the question of the stability of such dimerized solids in spin-1 systems, taking the anisotropic square lattice with bilinear and biquadratic spin-spin interactions as a paradigmatic model. The lattice can be viewed as a set of coupled spin-1 chains, which in the limit of vanishing inter-chain coupling are known to possess a stable dimer phase. 
We study this model using the density matrix renormalization group (DMRG) and infinite projected entangled-pair states (iPEPS) techniques, supplemented by the analytical mean-field  and linear flavor wave theory calculations. While the latter predicts the dimer phase to remain stable up to a reasonably large interchain-to-intrachain coupling ratio $r \lesssim 0.6$, the DMRG and iPEPS find that the dimer solid melts for much weaker interchain coupling not exceeding $r\lesssim 0.15$. We find the transition into a magnetically ordered state to be first order, manifested by a hysteresis and order parameter jump, 
%and finite correlation length, 
precluding the deconfined quantum critical scenario. The apparent lack of stability of dimerized phases in 2D spin-1 systems is indicative of strong quantum fluctuations that melt the dimer solid.
%We study the phase diagram of the coupled spin-1 bilinear-biquadratic Heisenberg chains model, and focus on the parameter regime where the decoupled chains have dimerized order. We use a combination of different numerical methods. We use the U(1)-symmetric density matrix renormalization group method to study long cylinders where the boundary effects are reduced by extrapolation technique. Also, we use the infinite projected-entangled-pair states method which by design deals with infinite systems to study the model in the thermodynamic limits. With increasing inter-chain couplings, both methods demonstrate that the dimerized order of 1D biliear-biquadratic chain melts very quickly into either N\'{e}el antiferromagnetic states or spin-nematic ferro-quadrupolar states, when the ratio of inter-chain couplings versus intra-chain couplings is smaller than $0.13$. The phase transitions are of first order as clear hysteresis is observed using both numerical methods. By relating our findings with a series of previous works, we argue that the spin-1 dimerized order is not competitive on two-dimensional lattices.
\end{abstract}

\maketitle

%\tableofcontents

%\section{\label{sec:intro}Introduction \protect}

Dimerized valence bond solid (DVBS) is a magnetic analog of an atomic crystal, with the spin singlets on the non-overlapping links of a lattice forming  a long-range order that spontaneously breaks the translation symmetry. Such DVBS are known to be the exact ground states of certain models~\cite{majumdar-ghosh,kumar-dimers2002} and are expected to appear naturally in spin-1/2 systems on bipartite lattices, where extensive numerical studies using density matrix renormalization group (DMRG), variational Monte Carlo (VMC) and tensor-network methods corroborate that VBS phases (including plaquette VBS) are stable on the square~\cite{gong-half-square2014,PhysRevLett.121.107202,PhysRevB.102.014417,PhysRevX.11.031034,LIU20221034}
%\cite{yu-half-square2012,gong-half-square2014} 
and honeycomb~\cite{gong-half-hon2013,ferrari-half-hon2017} lattices. Of particular interest is the effect of geometric frustrations that enhance the quantum fluctuations which can ``melt" the DVBS solid in favor of  the resonating valence bond (RVB) state~\cite{anderson-RVB,balents-review} -- a long sought-after quantum spin liquid in the paradigmatic $J_1-J_2$ model on the square lattice~\cite{gong-half-square2014,PhysRevLett.121.107202,PhysRevB.102.014417,PhysRevX.11.031034,LIU20221034}
%\cite{jiang-half-square2012,yu-half-square2012,gong-half-square2014}. 

While spin-1/2 models have been studied extensively, the appeal of higher-spin systems (where $S$ is not too large so as not to become quasi-classical) is that they allow for non-geometric frustration due to the nontrivial biquadratic interactions 
%of the type 
$(\mathbf{S}_i\cdot \mathbf{S}_j)^2$. The competition with the familiar Heisenberg term then results in a rich phase diagram that, in the case of two-dimensional (2D) $S=1$  model can potentially host more exotic phases, including the ferroquadrupolar and antiferroquadrupolar (spin nematic) orders~\cite{chub91,buch05,porr06,pcor17,pcor18, chan15,taol15}, as well as a putative quantum spin liquid that breaks the lattice point-group symmetry~\cite{niesen2017,hu2019}.
Experimentally, compounds like $\text{NiGa}\text{S}_2$ \cite{doi:10.1126/science.1114727, spin1exp1comp1, spin1exp2comp1} and $\text{Ba}_3\text{Ni}\text{Sb}_2\text{O}_9$ \cite{spin1exp3comp2,spin1exp4comp2} with $S=1$ moments have been proposed to be 
%closely related 
close to the spin-nematic phases. 
%And for special ratios of the biquadratic and bilinear couplings, 
Spin-1 model can also be fine-tuned to the SU(3)-symmetric points (there are two \cite{PhysRevLett.105.265301, PhysRevB.85.125116}) which lend themselves to possible realization in ultracold alkaline-earth atoms~\cite{Gorshkov_2010}.

This raises a question -- to what extent the DVBS phases, so central to the discussion of the dimer models~\cite{rokhsar-kivelson,moessner-sondhi} and RVB spin liquids~\cite{anderson-RVB,balents-review},
are prevalent in higher spin models, in particular in spin-1 counterparts? In principle, spin singlet state is allowed to form on a pair of sites for an arbitrary spin representation of SU(2), so there is no fundamental obstruction. Yet numerical studies point to the lack of stability of DVBS states in SO(3)-symmetric spin-1 models on either the square or the honeycomb lattices in 2D~\cite{pcor17, spin1j1j2squa, spin1bbhhon, spin1j1j2hon}.
This is to be contrasted with 1D spin chains, where the DVBS state (sometimes called spin-Peierls state) is well documented in the bilinear-biquadratic spin-1 model~\cite{lauc06,pixley-spin1}. 
To understand the reason for this dichotomy, we consider the 
%most general SO(3)-symmetric 
paradigmatic bilinear-biquadratic 
spin-1 model on an anisotropic square lattice with the vertical links weaker than the horizontal ones: 
\begin{align}
    H =  &\sum_{i,j} \bigg[J_\parallel\boldsymbol{S}_{i,j} \cdot \boldsymbol{S}_{i+1,j} + K_\parallel (\boldsymbol{S}_{i,j} \cdot \boldsymbol{S}_{i+1,j})^2\bigg] \notag\\ & +
    \sum_{i, j} \bigg[J_\perp\boldsymbol{S}_{i, j} \cdot \boldsymbol{S}_{i, j+1} + K_\perp (\boldsymbol{S}_{i, j} \cdot \boldsymbol{S}_{i,j+1})^2\bigg].
    \label{eqn:BBH}
\end{align} 
The degree of the lattice anisotropy can be captured by the ratio 
$r\equiv J_\perp/J_\parallel=K_\perp/K_\parallel$ 
(we assume the last equality for simplicity, but it does not affect our conclusions qualitatively).
Clearly, $r=0$ corresponds to the limit of decoupled spin chains, for which we know the DVBS phase to be stable as long as $K_{\parallel}$ is negative and $|K_{\parallel}| > |J_\parallel|$~\cite{lauc06,pixley-spin1}.
How far does this DVBS phase extend toward the isotropic square lattice limit ($r=1$)? 

\begin{figure}[thb]
    \centering
    \includegraphics[scale=0.44]{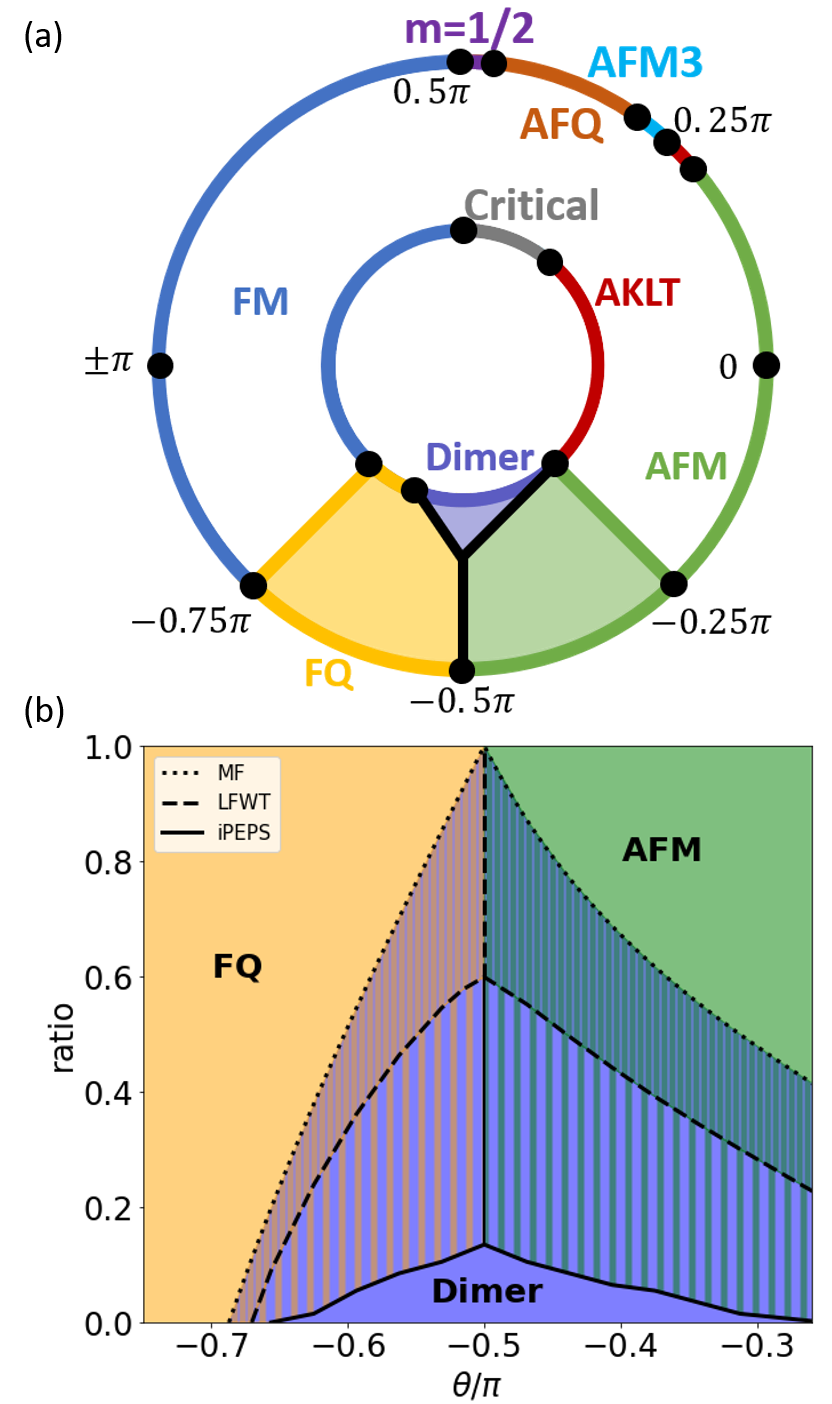}
    \caption{\footnotesize{(a) A schematic illustration of the piled-up phase diagrams for the spin-1 BBH chain (inner circle) and the spin-1 BBH model on the square lattice (outer circle), with the shaded region in-between representing the phase diagram of the coupled spin-1 BBH chains. The radical axis describes the relative strength of inter-chain couplings. (b) Phase diagram for the coupled spin-1 BBH chains, where mean-field, the linear flavor wave theory and iPEPS results are plotted. The solid lines indicate the actual phase boundaries, which are of first order.}}
    \label{fig:phase_diagrams}  
\end{figure}
% \begin{figure}[H]
%     \centering
%     \includegraphics[scale=0.55]{plots/phase_diagrams_4.png}
%     \caption{\footnotesize{(a) Mean-field phase diagram obtained using classical ansatz, $\left|\text{DM}\right>$, $\left|\text{FQ}\right>$ and $\left|\text{NAF}\right>$. The linear flavor wave theory correction is also considered.  (b) Phase diagram for the coupled spin-1 BBH chains. The errorbar indicates the range of hysteresis.}}
%     \label{fig:phase_diagrams}  
% \end{figure}

To answer this question, in this Letter we have performed numerical analysis by two different tensor network methods, DMRG and infinite projected entangled-pair state (iPEPS), supplemented by the analytical mean-field (MF) and linear flavor-wave theory (LFWT) calculations. Our main findings are summarized in Fig.~\ref{fig:phase_diagrams}, utilizing the commonly used parametrization $J_\parallel=\cos\theta$, $K_\parallel=\sin\theta$ in terms of the polar angle $\theta$. We find that a relatively weak interchain coupling ($r_c \lesssim 0.15$) is sufficient to melt the VBS solid, yielding one of the rotational-symmetry-broken long range ordered phases: either the N\'eel (at $\theta > -\pi/2$) or the ferroquadrupolar phase (at $-2\pi/3 \lesssim \theta <\! -\pi/2$). Since these phases break a different symmetry then the SO(3)-symmetric VBS state, a direct second-order phase transition between such states is seemingly not allowed in the 
%conventional 
Landau--Ginzburg paradigm. However, when topological defects of either order parameter are considered, a field theory supporting a putative continuous phase transition between the VBS and N\'eel state can be written down, oultining the framework of so-called deconfined quantum criticality~\cite{senthil-dqcp-2004,dqcp-prb2004}. This 
%consideration 
provides a second motivation for the present study, namely: can the model in Eq.~(\ref{eqn:BBH}) support such a continuous phase transition? Our results show that the answer is negative, with a clear hysteresis indicating a first-order phase transition from the VBS into either the N\'eel or FQ ordered state.

\noindent
\textit{DVBS phase.} In 1D, the existence of the dimerized VBS phase is well established, with the simplest wavefunction being a product of dimers on alternating bonds along the decoupled chains:

\vspace{-4mm}\begin{equation}
    |\text{DVBS}\rangle = \prod_j\prod_{i=2m+1} |\text{dimer}(S_{i,j},S_{i+1,j})\rangle, \label{eq:VBS}
\end{equation}
bearing in mind that the singlet state of two spins-1 is written as $|\text{dimer}(S_1,S_2)\rangle = \frac{1}{\sqrt{3}}(|1\bar{1}\rangle +|\overline{1}1\rangle-|00\rangle)$. 
The corresponding dimer order parameter $D_h$ is defined by

\vspace{-4mm}\begin{equation}
    \langle D_h\rangle = \langle \mathbf{S}_{i,j}\cdot\mathbf{S}_{i+1,j} \rangle - \langle \mathbf{S}_{i+1,j}\cdot\mathbf{S}_{i+2,j} \rangle.\label{eq:Dh}
\end{equation}
Note that the  $|\text{DVBS}\rangle$  ansatz is a zero-length entangled state with no correlation between the dimers, achieving the maximal possible $|\langle D_h\rangle| = S(S+1)=2$. A most general dimer state need not be a product state, and correlations can develop between the dimers; in all cases a non-vanishing Landau order parameter (\ref{eq:Dh})  serves as a definition of the DVBS long-range order~\footnote{In this work, we shall not consider AKLT-like states for spin-1, which are confusingly sometimes also called VBS states. For the remainder of this work, we will use the term VBS to refer to the translational symmetry broken, dimerized states, sometimes also called spin-Peierls states}. 
%We caution the reader that some authors use the term `VBS' to refer to the AKLT-like states that do not break the translational symmetry along the chain direction, this is not the definition we adopt in this work.

%\anc{Yi: the way I define the $D_h$ parameter in Eq.~(\ref{eq:Dh}) means it cannot exceed 2, but your plot in Fig. 3a suggests that you have a different convention with an extra 1/2? Also, note that you define a different quantity $D_h$ in the supplementary (a single dimer rather than difference between two nearby ones). We must be consistent. One way is to define $D_h=D(r,r+x)-D(r+x,r+2x)$ in terms of a single dimer operator $D(r_1,r_2)$. Then, you just need to change your Eq.~(S2) slightly.} \yic{The definition here is the correct one. There was a typo in E1.~(S2), which has been resolved. And $D_h$ is 2 for a perfect dimer state. In our case, its value in the dimer phase happens to be near 1.} \anc{ok, got it, thanks}

\vspace{1mm}
\noindent
\textit{Competing phases.} In the 2D limit of isotropic square lattice ($r=1$), prior iPEPS results find no DVBS phase~\cite{pcor17}, instead the interval $\theta \in[-3\pi/4,-\pi/4]$ is occupied by the FQ (yellow) and N\'eel (green) phases, as indicated in Fig.~\ref{fig:phase_diagrams}a. The N\'eel phase %needs no explanation, it 
is characterized by the staggered magnetization $m_s$. The FQ phase, on the other hand, 
%is characterized by
features a vanishing magnetic moment $\langle \mathbf{S}\rangle=\mathbf{0}$, but breaks the SO(3) spin symmetry in a more subtle way.
%, visualized as rotations of the $|S^z=0\rangle$ state. To see this, 
Namely, FQ phase can be written (up to an overall phase) as a linear superposition with \textit{real} coefficients of three quadrupolar basis states $|x\rangle=\frac{i}{\sqrt{2}}(|1\rangle -|\bar{1}\rangle)$,  $|y\rangle=\frac{1}{\sqrt{2}}(|1\rangle +|\bar{1}\rangle)$, $z=-i|0\rangle$~\cite{Penc2011}:
\vspace{-1mm}
\begin{equation}
    |FQ\rangle = d_x |x\rangle + d_y|y\rangle + d_z|z\rangle.\label{eq:FQ}
\end{equation}
The real coefficients $d_\alpha$ form the components of a director, which spontaneosly breaks the SO(3) symmetry. One can show that the FQ state has a non-zero value of the ferroquandrupolar operators defined as a traceless symmetric tensor $\hat{Q}^{\alpha\beta}\equiv \hat{S}^{\alpha}\hat{S}^{\beta} + \hat{S}^{\beta} \hat{S}^\alpha - \frac{2}{3} S(S+1)\delta_{\alpha \beta}$. One way to define the FQ order parameter is proportional to the trace of this matrix squared: $II_Q = -\frac{1}{2}\mathrm{Tr}(\langle\hat{Q}\rangle^2)$~\cite{pcor17}.

\vspace{1mm}
\noindent
\textit{Mean-field and LFWT insights.}
Using the product state ans\"atze for the  N\'eel, FQ and DVBS phases, we obtain the  expressions for the mean-field energies (see SM):
\begin{align}
    E_{\text{N}}^{\text{mf}}(r) &= -(1+r)\cos{\theta}+(S^2+1)(1+r)\sin{\theta} \nonumber\\
    E_{\text{FQ}}^{\text{mf}}(r) &= \big(S^2+\frac{1}{2}+\frac{1}{2}\Theta(\frac{5}{4}-S)\big)(1+r)\sin{\theta} \label{eqn:mf}\\
     E_{\text{DVBS}}^{\text{mf}}(r) &= -\frac{S+1}{2S}\cos{\theta}+\frac{(S+1)^2}{3}(2+r)\sin{\theta}, \nonumber 
\end{align}
which for $S=1$ results in the transitions indicated by the dotted lines in Fig.~\ref{fig:phase_diagrams}b. The mean-field treatment overestimates the stability of the DVBS phase and we use the linear flavour-wave theory to account for the quantum fluctuations around the FQ and N\'eel phase, respectively. Starting from the N\'eel phase, the LFWT treatment is equivalent to the Holstein--Primakoff bosons, with the resulting energy lowered by the zero-point fluctuations (see SM). The flavor-wave expansion around the FQ order requires more care, 
%with the principal idea of 
representing the spin operators $S^\alpha$ and quadrupolar operators $Q^{\alpha\beta}$ in terms of the bilinears of the three flavors of Schwinger bosons forming a fundamental representation of group SU(3), namely: $S^\alpha = -i\varepsilon_{\alpha\beta\gamma}b^\dagger_\beta b_\gamma,\; Q^{\alpha\beta} = \frac{2}{3}\delta_{\alpha\beta} - b_\alpha^\dagger b_\beta - b_\beta^\dagger b_\alpha$. One then treats the FQ phase as a Bose--Einstein condensation of one of the boson flavors $\langle b_z \rangle \neq 0$, and uses the  
%Hilbert space 
constraint $\sum_\alpha b^\dagger_\alpha b_\alpha = 1$ to express the condensate fraction as 
\begin{equation}
    \langle b_z \rangle  = \sqrt{1-b^\dagger_x b_x-b^\dagger_y b_y},
\end{equation}
where expanding the square root to first order in boson bilinears amounts to LFWT. Diagonalizing the resulting Hamiltonian and computing the zero-point energy correction to the mean-field expression (\ref{eqn:mf}) makes the FQ and N\'eel phases more energetically favourable, thus lowering the critical value $r_c$ at which the VBS phase melts, shown by the dashed line in Fig.~\ref{fig:phase_diagrams}b. 
%For VBS/N\'eel transition, we find $0.25 \lesssim r_c^{N} \lesssim 0.6$ from LFWT. 

\vspace{1mm}
\noindent 
\textit{iPEPS.} In order to provide an unbiased corroboration of the above LFWT results, we have used the iPEPS method~\cite{PhysRevLett.101.250602}, which has been  previously demonstrated to describe well both the FQ and N\'eel order of spin-1 on the square lattice~\cite{pcor17}. In iPEPS, the wavefunction on the (infinite) square lattice is written in terms of a product of tensors, with the contraction over the auxiliary indices that are defined on the links of the lattice. 
Based on the known phases in 1D and 2D, we choose the bipartite tiling by a $2\times 1$ unit cell containing two distinct tensors $a$ and $b$ which we then variationally optimize~\cite{liao2019}. Such ansatz can describe the staggered pattern of the N\'{e}el ordered states as well as the DVBS covering
\begin{equation}\label{eq:ansatz}
|\psi(a,b)\rangle = \vcenter{\hbox{\includegraphics[scale=0.25]{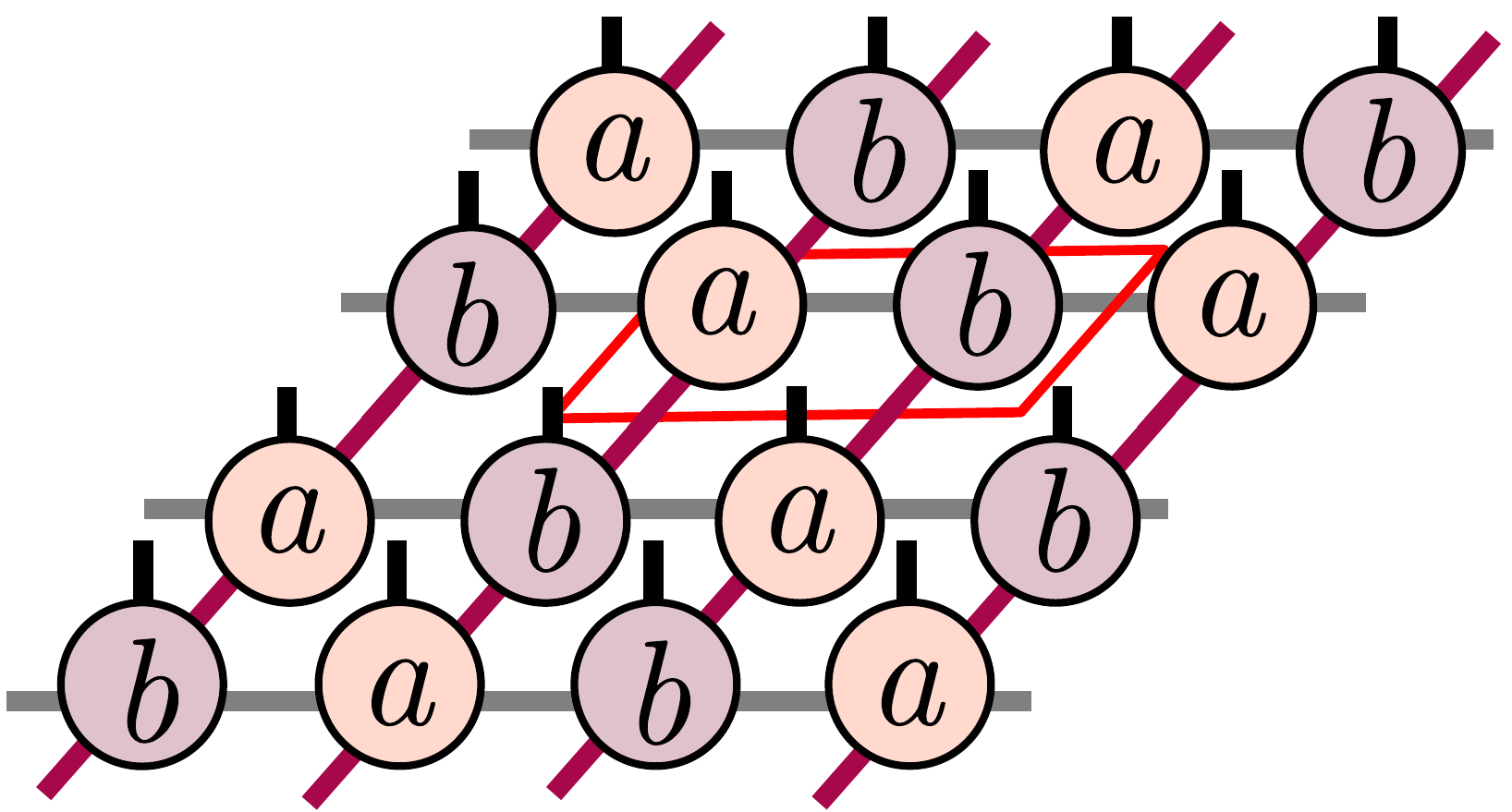}}}.
\end{equation}
The physical indices associated to physical spin $S=1$ degrees of freedom are represented by black vertical lines. The approximation is controlled by the bond dimensions of auxiliary indices in horizontal (gray indices $l,r$) and vertical direction (magenta $u,d$), now denoted $D_x$ and $D_y$ respectively, which need not to be the same. Indeed, the decoupled chain limit $r\to0$ can be described just with $D_y=1$. Evaluation of energy and order parameters requires additional \textit{environment} tensors whose precision is controlled by their  environment dimension $\chi$, here constructed by corner transfer matrix method~\cite{corboz2014}.  In our simulations, we found it sufficient to set $D_x=9$, $D_y=4$ and $\chi=16$ in order to obtain well converged results in all three phases ($D_x=9$ turns out to be the minimal bond dimension that allows to realize a variationally competitive SU(2)-symmetric DVBS state in the limit of decoupled chains). To verify, we have run the iPEPS calculations with more exacting parameters ($D_x=9$, $D_y=6$, and $\chi$ up to $32$)  at several points in each phase, which gave nearly identical results. The relatively small environment dimension $\chi$ necessary to obtain converged results points to the short correlation lengths of optimized iPEPS in the competing ordered phases. We refer the reader to the Supplementary Materials for further details of the iPEPS simulations.

\begin{figure}[tb]
    \flushleft
    \includegraphics[scale=0.32]{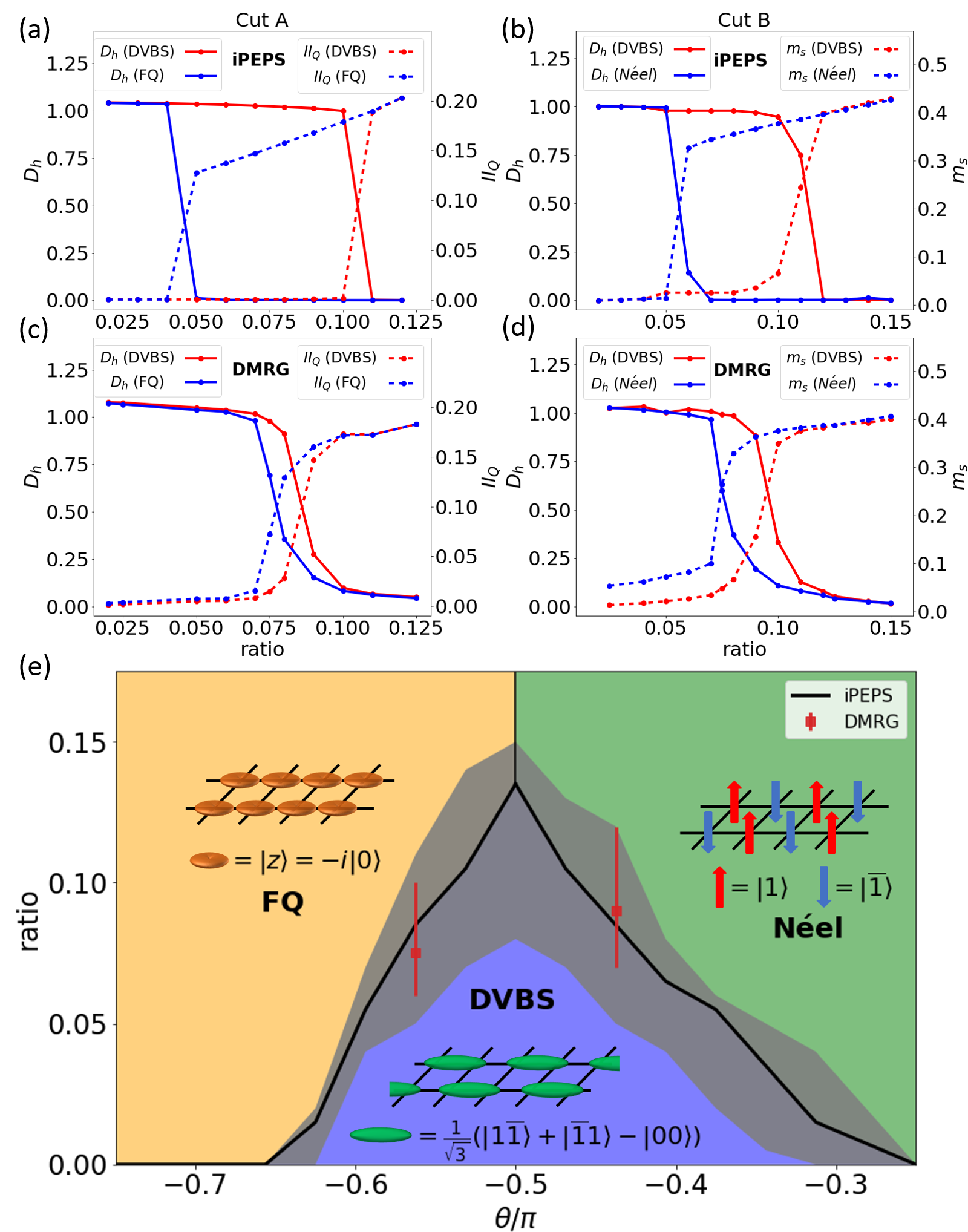}
    \caption{\footnotesize{Hysteresis.  (a)-(d) Order parameters for cut A ($\theta=-0.5625\pi$) and B ($\theta=-0.4375\pi$). States for each ratio are optimized using both dimerized initial states (blue curves) and ferro-quadrupolar states (red curves). (e) Hysteresis is indicated by shaded region (iPEPS) and errorbars (DMRG).}}
    %\hspace{10in}
    \label{fig:hysteresis}
\end{figure}

The resulting phase diagram is shown in Fig.~\ref{fig:phase_diagrams}b and Fig.~\ref{fig:hysteresis}e, with the solid black line indicating the boundary of the DVBS phase. It is immediately apparent that a small value of interchain coupling (anisotropy ratio $r_c\lesssim 0.15$) is sufficient to melt the DVBS phase in favor of a long-range ordered magnetic or nematic (N\'eel or FQ) phase. This is completely consistent with the prior iPEPS results on the isotropic square lattice ($r=1$) where the DVBS order is conspicuously absent~\cite{pcor17}.

%\anc{Too detailed to mention this here? Moved to SM...}\\
%\an{We also checked that the bipartite tiling (staggered dimer state) and the stripe tiling (dimer state) give very close energies in the limit of $r\to 0$. For example, at $(\theta, \text{r})=(-0.5625\pi, 0.02)$, with $D_x=9$, $D_y=4$ and $\chi=16$, $E_{\text{bipartite}}=-2.558822$, $E_{\text{stripe}}=-2.558595$. }

%Among various phases of the $S=1$ BBH model defined on different lattices, there is one unique phase that is only found in the one-dimensional chain - a dimer phase\cite{lauc06}. The dimer phase preserves the SU(2) spin-rotational symmetry yet breaks the lattice translational symmetry, which makes it quite different from other phases in the phase diagram. In the phase diagrams of the $S=1$ BBH model on 2D lattices, however, the dimer phase is always replaced by either the N\'{e}el AFM phase or the ferro-quadrupolar (FQ) phase\cite{pcor17, pcor18, chan15, taol15}. This brings about a natural question - is the dimerized order stable for $S=1$ Heisenberg-type or even more general models on 2D lattices? In the meantime, it is worth mentioning that the N\'{e}el AFM and FQ phases both break the spin-rotational symmetry while preserving the $C_4$ lattice rotation symmetry (the lattice translation symmetry can be either partially broken or preserved). Hence, based on the Landau's paradigm, there should exist some dimer "melting" transitions if one can construct a model that adiabatically connects the 1D and 2D phase diagrams, as shown by \ref{fig:schem_phase_diag}.

\vspace{1mm}
\noindent
\textit{DMRG.} As a complementary method, we also use the density matrix renormalization group method to simulate the coupled chains. DMRG is a numerical method first proposed to study 1D systems~\cite{white92} and subsequently generalized to the studies of finite 2D systems in a cylindrical geometry~\cite{dmrg2d}. 
%It performs optimization of Matrix product state wave function $|\psi(\{M\})\rangle=\sum_{\{s\}}Tr[M_1^{s_1}M_2^{s_2}\ldots M_N^{s_N}]|\{s\}\rangle$, parametrized by rank-3 on-site tensors $\{M^s_{lr}\}$ associated with physical sites~\cite{ostlund1995,perez2007}. Similarly to iPEPS, $s$ is a physical index and $l,r$ are auxiliary indices of left and right bond of site.
Aligning the chains along the cylinder direction ($x$) with open boundary conditions, we chose long cylinders of dimension $N_x\times N_y$ with $N_x=30-80$ to minimize the effect of the boundaries. The results were then extrapolated to the 1D thermodynamic limit $N_x \to \infty$. The computational complexity grows exponentially with the cylinder circumference, which was therefore kept at $N_y=4$ for most of the calculations. We checked that larger $N_y=6$ simulations %for small $N_x(=30)$, which
do not qualitatively change our conclusion. We used DMRG implementation provided by ITensor library~\cite{itensor}, with explicit U(1) symmetry.
The maximum number of states kept in our simulations were $M=1600$, achieving acceptable truncation errors 
%on the density matrices
less than $8\times10^{-5}$. 

\begin{figure}[tb]
    \centering
    \includegraphics[width=0.33\textwidth]{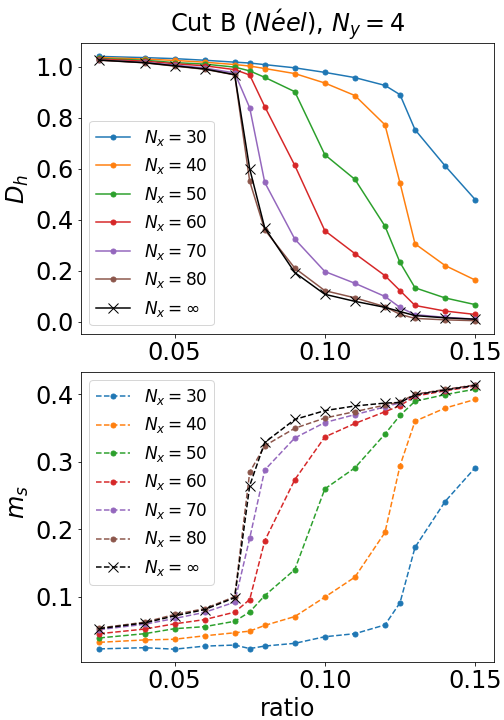}
    \caption{\footnotesize{Finite size scaling of the VBS order parameter $D_h$ and the staggered magnetization $m_s$ in DMRG calculations, plotted along the vertical cut B ($\theta=-0.4375\pi$) in the phase diagram (Fig.~\ref{fig:hysteresis}e). Different symbols/colors label the cylinder lengths $N_x$, extrapolated to the thermodynamic limit. The N\'{e}el AFM state was used to initialize the density matrices.}}
    \label{fig:finite_size_scaling}
\end{figure}

It turns out that cylinders of $N_y=4$ are able to capture the phase transitions as the DVBS order melts very quickly even with weak inter-chain couplings, corroborating our iPEPS results. Shown in Fig.~\ref{fig:finite_size_scaling} are the two orders parameters along the DVBS--N\'eel phase transition, averaged over the middle portion $\Delta N_x\times  N_y=12\times 4$ of the cylinders for all $N_x$ so as to avoid the boundary effects. We note in passing that care must be taken in computing the order parameter in DMRG, since spontaneous breaking of the SO(3) spin symmetry is only possible in the thermodynamic limit. To overcome this, we introduced the pinning fields on the open boundaries of the cylinder (see SM for more details), as is the standard practice in U(1)-symmetric DMRG calculations.

It is clear from the order parameter plots Fig.~\ref{fig:finite_size_scaling} that 
there is a (narrow) regime of coexistence, indicating that the DVBS--N\'eel transition is first order. To further corroborate this finding, we analyze below the hysteresis of both this and the DVBS--FQ transition.

\vspace{1mm}
\noindent
\textit{Hysteretic melting of the DVBS order.} We corroborate the first-order nature of the phase transitions from DVBS phase by observing the hysteresis. Indeed, we found that the critical value $r_c$ at which the DVBS phase melts depends on the choice of the initialization  
%of the density matrix 
in DMRG (or the initial choice of tensors in iPEPS). To systematically investigate this effect, we have prepared the initial state to either be the $|\text{DVBS}\rangle$ product state in Eq.~\eqref{eq:VBS} or one of the ordered states: FQ for $\theta<-\pi/2$ (``cut A" in Fig.~\ref{fig:hysteresis}) or N\'eel for $\theta>-\pi/2$ (``cut B" in Fig.~\ref{fig:hysteresis}). We then plot the dimer order parameter $\langle D_h\rangle$ (solid lines in Fig.~\ref{fig:hysteresis}), which shows a clear hysteresis loop traced between the blue and red lines in Figs.~\ref{fig:hysteresis}a-\ref{fig:hysteresis}d depending on the choice of the initial condition. We similarly plot with the dashed lines the FQ order parameter ($II_Q$, defined below Eq.~\ref{eq:FQ}) along cut A in Figs.~\ref{fig:hysteresis}(a,c) and the
staggered magnetization $m_s$ along cut B in Figs.~\ref{fig:hysteresis}(b,d). In all cases, a clear hysteresis associated with a discontinuous phase transition is observed. The width of the hysteresis is narrower in our DMRG data compared with iPEPS, which we attribute to the practical limitations on the circumference of the cylinder used in DMRG calculations.

We determine the position of the DVBS melting phase transition by following 
the energy level crossing between the two differently initialized configurations (see SM). The transition $r_c$ thus obtained falls roughly halfway inside the hysteresis loop (greyed area in the phase diagram Fig.~\ref{fig:hysteresis}e) -- this is how the phase boundaries (solid black lines) were obtained in Figs.~\ref{fig:phase_diagrams}b and \ref{fig:hysteresis}e. We have also verified that the correlation functions all remain short-ranged across the DVBS melting transition (see SM), again corroborating the discontinuous nature of the transition.

\vspace{1mm}
\noindent
\textit{Discussion.} 
As the comparison between the MF results, LFWT and DMRG/iPEPS indicates, quantum fluctuations tend to destabilize the dimerized VBS order even at very weak interchain couplings $r_c\lesssim 0.15$. We find that the DVBS state, even when stable, is not a simple product state such as in the MF ansatz \eqref{eq:VBS}, but that dimer-dimer correlations are pronounced, eventually melting the dimer order. 

While the present study focuses on the frustration induced by the biquadratic interaction, the implications hold for more general models with geometric frustrations. For instance, no VBS state appears to be stable on $J_1$--$J_2$ spin-1 square lattice~\cite{spin1j1j2squa}. In fact the only isotropic lattice (that the authors are aware of) appearing to support a VBS state is the honeycomb  $J_1$--$J_2$ model, where DMRG calculations find a stable plaquette VBS order~\cite{spin1j1j2hon}.
%It is intriguing that the $S=1$ DVBS order does not appear to  be stable in any isotropic 2D lattices that the authors are aware of, with the exception perhaps of the $J_1-J_2$ model on a honeycomb lattice, where a plaquette VBS state was found in DMRG calculations~\cite{spin1j1j2hon}. By contrast, The present study helps shed light on the reasons for this instability due to the quantum entanglement between neighboring chains. 

%% DO we need this remark? 
%In hindsight, we may glean the reason for why the VBS state might be more stable on a honeycomb lattice -- it has a smaller coordination number 3, meaning that each dimerized chain is connected to at most one other.  This in turn implies that interchain hybridization plays a lesser role than on the square lattice, thus making the VBS phase more stable.

The discontinuous nature of the DVBS quantum melting transition we found precludes the possibility of a deconfined quantum criticality mentioned earlier in the context of spin-$1/2$ models, in agreement with the finding in Ref.~\onlinecite{neelvbs-spin1-1st} for a more complicated spin-1 model. It is conceivable that the deep reason for the difference in the stability of the VBS phases between spin-1/2 and spin-1 models lies in the Lieb--Schultz--Mattis--Oshikawa--Hastings theorem~\cite{LSM-theorem,oshikawa-LSM,hastings-LSM}, which prohibits existence of gapped non-topological phases in spin-1/2 systems unless the unit cell is doubled due to the dimer formation. There is no such restriction for integer spin systems however, with a featureless correlated paramagnet state possible, which has been suggested in several numerical studies on spin-1 models~\cite{pcor17,pcor18,spin1bbhhon,chan15}. As the present work indicates, spin-1 models offer a rich palette of quantum phases that deserve future investigation.

\begin{acknowledgments}
We thank Miles Stoudenmire  for stimulating discussions and give credit to Shuyi Li for the guidance in the LFWT analysis. Y.X. and A.H.N. acknowledge the support of the National Science Foundation Division of Materials Research under the Award DMR-1917511. T.F. was supported by the Welch Foundation grant no. C-1818. J.H. was supported by the European Research Council (ERC) under the European Union’s Horizon 2020 research and innovation programme (grant agreement No 101001604). The majority of the calculations were performed on the clusters supported by the Big-Data Private-Cloud Research Cyberinfrastructure MRI-award funded by NSF under grant CNS-1338099 and by Rice University's Center for Research Computing (CRC). 
\end{acknowledgments}

% The \nocite command causes all entries in a bibliography to be printed out
% whether or not they are actually referenced in the text. This is appropriate
% for the sample file to show the different styles of references, but authors
% most likely will not want to use it.
% \nocite{*}

\renewcommand{\emph}[1]{\textit{#1}}
\bibliography{spin1,spin-half}% Produces the bibliography via BibTeX.

\clearpage
\setcounter{equation}{0}
\setcounter{figure}{0}
\setcounter{table}{0}
\makeatletter
\renewcommand{\theequation}{S\arabic{equation}}
\renewcommand{\thefigure}{S\arabic{figure}}
\renewcommand{\bibnumfmt}[1]{[#1]}
\renewcommand{\citenumfont}[1]{#1}

%%%%%%%%%% Merge with supplemental materials %%%%%%%%%%
\begin{widetext}
\begin{center}
	\Large{Supplementary Materials for ``\thistitle"}
\end{center}
\end{widetext}

\section{\label{appendix:methods_peps} Infinite Projected Entangled-Pair States (iPEPS)}

The ground-state wave function  is parametrized by a set of rank-5 on-site tensors $\{a^s_{uldr}\}$ associated with physical sites in the unit cell (depending on the spatial pattern) which then tiles the entire square lattice. The physical index $s$ runs over states of $S=1$ degrees of freedom while $u,l,d$, and $r$ are auxiliary indices of \textit{bond dimension} $D$ associated with up, left, down and right bonds of each site on the square lattice. 
The physical wavefunction on (infinite) square lattice is then formally  obtained by contracting on-site tensors along pairs of auxiliary indices common to the bonds of the lattice 
%$|\psi(\{a\})\rangle=\sum_{\{s\}}Tr_{aux}[\ldots a^{s_{-1}}a^{s_0}a^{s_1} \ldots]|\{s\}\rangle$.
\begin{equation}
|\psi(\{a\})\rangle=\sum_{\{s\}}\mathrm{Tr}[\ldots a^{s_{-1}}a^{s_0}a^{s_1} \ldots]\;|\dots s_{-1}\, s_0\, s_1 \ldots\rangle
\end{equation}
%\anc{Yi: why is there summation over s? The wavefunction on the l.h.s. should depend explicitly on the physical indices $s_i$, not be to be summed.}

The quality of the approximation is controlled by $D$, starting from trivial $D=1$ iPEPS describing only mean-field wave functions to highly-entangled states as $D$ grows. Tractable manipulation of iPEPS
is realized by considering only reduced density matrices of finite subsystems, which are sufficient to compute physical quantities of interests such as energy, local order parameters or correlation functions. We construct such reduced density matrices from iPEPS environments which provide finite-dimensional embedding of subsystems.   The environments are obtained by the corner transfer matrix renormalization group (CTMRG) algorithm~\cite{ctmrg98,corboz2014}, with their precision controlled by \textit{environment} bond dimension $\chi$. For given iPEPS, one recovers its exact reduced density matrices in the limit of $\chi\to\infty$.
To find optimal values for elements of on-site tensors, we perform their gradient-based optimization using automatic differentiation~\cite{liao2019}. These simulations were realized using \textit{peps-torch} library~~\cite{pepstorch}.

The variational energy per site $e$ of iPEPS state given in Eq.~\ref{eq:ansatz} is obtained by evaluating four non-equivalent contributions
\begin{equation}
\begin{split}
e= &\sum_{i=0,1} \langle J_\parallel\boldsymbol{S}_{i,0} \cdot \boldsymbol{S}_{i+1,0} + K_\parallel (\boldsymbol{S}_{i,0} \cdot \boldsymbol{S}_{i+1,0})^2 \rangle + \\
&\sum_{j=0,1} \langle J_\perp\boldsymbol{S}_{0, j} \cdot \boldsymbol{S}_{0, j+1} + K_\perp (\boldsymbol{S}_{0, j} \cdot \boldsymbol{S}_{0,j+1})^2\rangle.
\end{split}
\end{equation}
To evaluate the above terms, we construct four distinct reduced density matrices (RDMs) of nearest-neighbor sites using environment tensors $\{C,T\}$ obtained from CTMRG. Assuming tensor $a$ is assigned to site $[i,j]=[0,0]$, the first two RDMs  are $\rho_{2\times1}([0,0], [1,0])$, $\rho_{1\times2}([0,1], [0,2])$
\begin{equation}
\vcenter{\hbox{\includegraphics[scale=0.33]{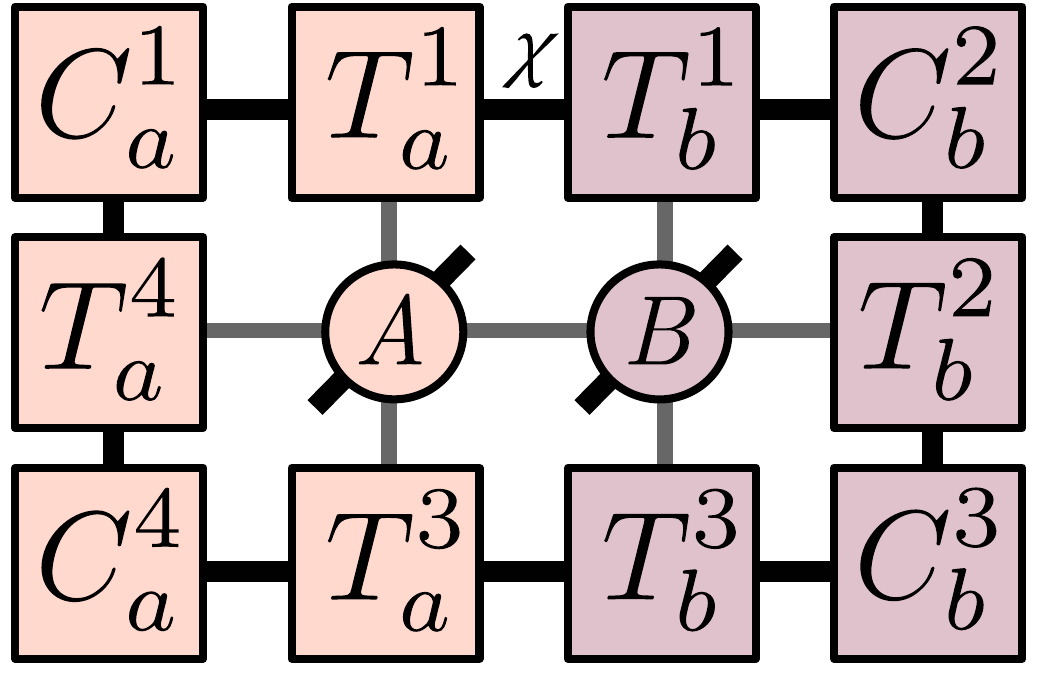}
}},
\vcenter{\hbox{\includegraphics[scale=0.33]{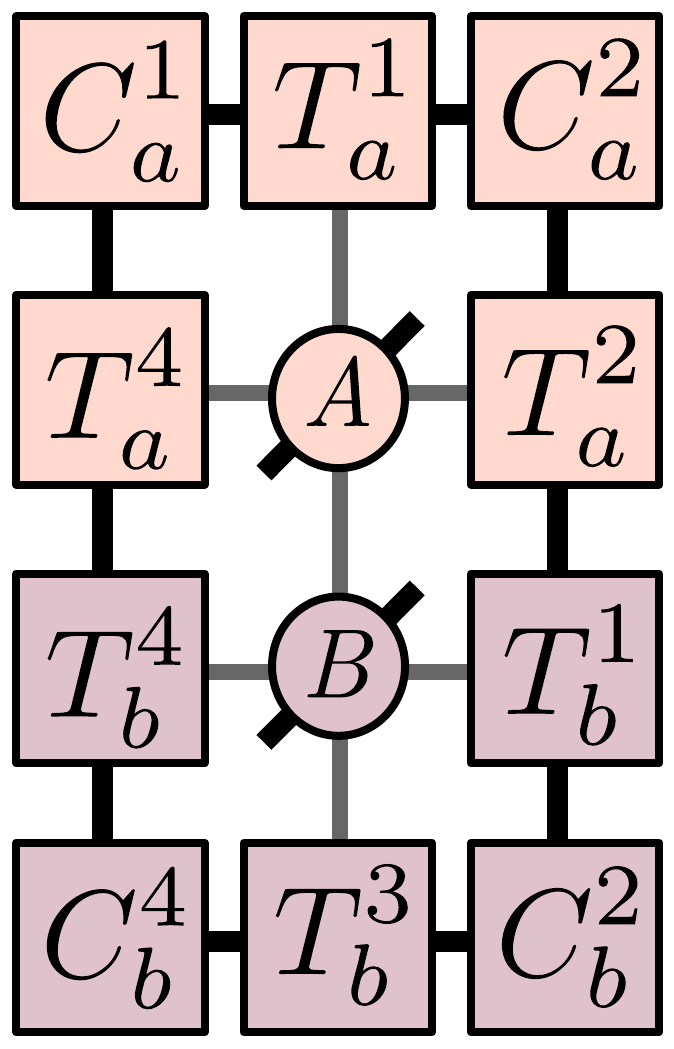}}},\ldots
\end{equation}
and the remaining two, $\rho_{2\times1}([1,0], [2,0])$ and $\rho_{1\times2}([0,0], [0,1])$, which we do not show can be constructed analogously. Tensors $A$ and $B$ are double-layer tensors obtained by contracting the physical index, i.e., $A_{(uu')(ll')(dd')(rr')}=\sum_s a^s_{uldr} {a^*}^s_{u'l'd'r'}$,
while sites with open physical indices, $A^{ss'}_{(uu')(ll')(dd')(rr')}= a^s_{uldr} {a^*}^s_{u'l'd'r'}$, have diagonal black lines. 

The optimization is done by repeating four steps: 
(i) perform CTMRG for state $|\psi(a,b)\rangle$ and obtain environment tensors $\{C,T\}$, (ii) evaluate variational energy $e$, (iii) compute gradients $g_a=\partial e/\partial a$, $g_b=\partial e/\partial b$, (iv) update tensors
$a$ and $b$ using L-BFGS method with a backtracking line search. 
The optimization terminates once the relative difference in variational energies between consecutive steps becomes lower than $10^{-8}$. 
For details regarding L-BFGS gradient descent, which is a quasi-Newton method constructing an approximate Hessian from past gradients, and line search see Ref.~\cite{numerical}.

In the main text, we have used exclusively bipartite tiling of the square lattice with $2\times1$ unit cell. In the dimerized phase this choice leads to \textit{staggered} dimer state (sVBS). The second option is to use stripe tiling, where horizontal dimers are aligned in columns (cVBS). We have verified, that these two choices are very close in energies in the dimer phase. For example, at $(\theta, \text{r})=(-0.5625\pi, 0.02)$, with $D_x=9$, $D_y=4$ and $\chi=16$, $e_{\text{sVBS}}=-2.558822$, $E_{\text{cVBS}}=-2.558595$. 

\begin{figure}[tb]
    \flushleft
    \includegraphics[scale=0.4]{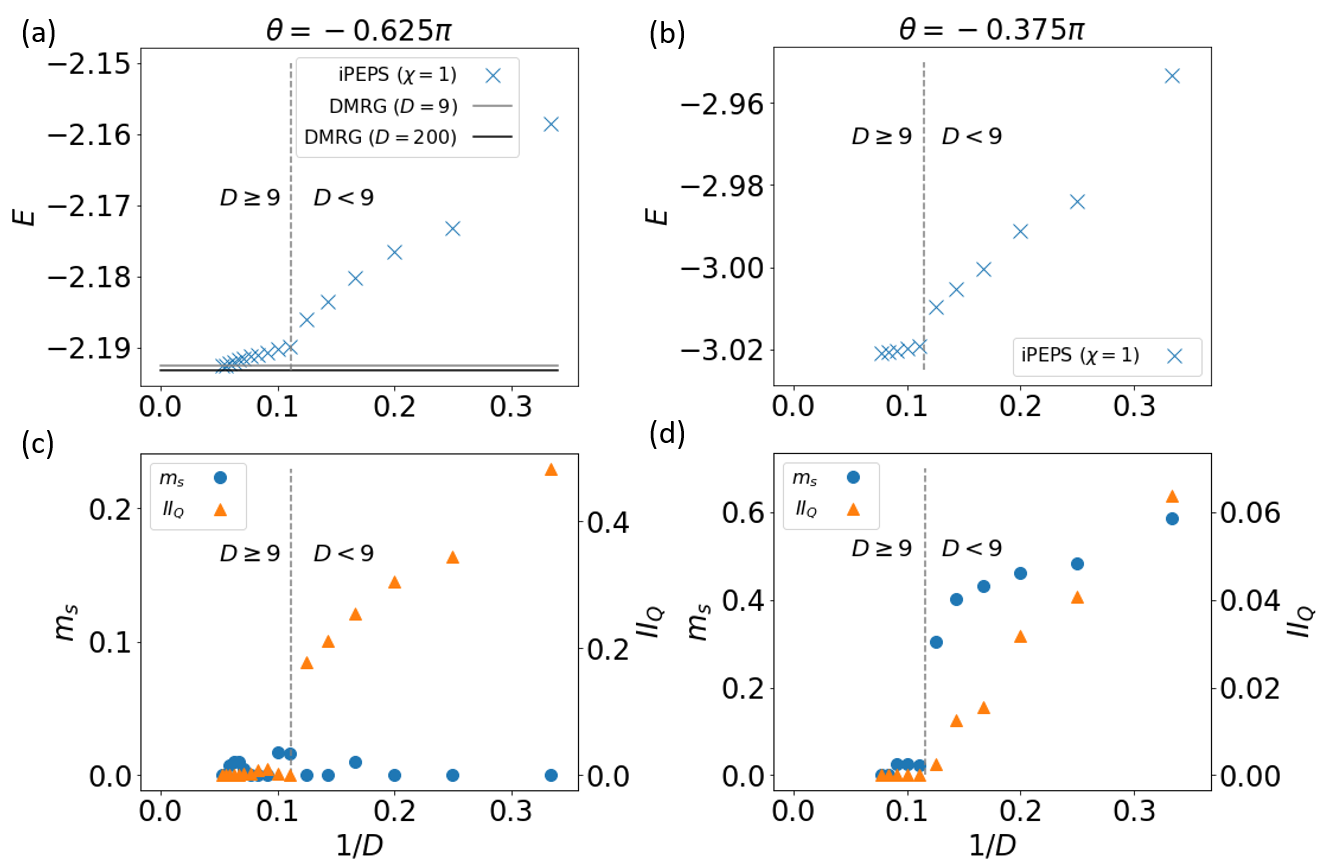}
    \caption{\footnotesize{Spurious SU(2)-symmetry breaking induced by  finite-$D$ effects in dimerized phase of decoupled chains, $r=0$, for $\theta=-5/8\pi$ (left) and $\theta=-3/8\pi$ (right). Keeping $D_y=1$, for $D_x\ge9$ all order parameters $m_s$ and $II_Q$ become small.}}
    \label{fig:dx1}
\end{figure}

Although the dimerized phase of coupled spin-$1$ chains is gapped,
its accurate description in terms of iPEPS requires a relatively large bond dimension. We study the limiting case of iPEPS ansatz with $D_y=1$, which in effect parametrizes the ground state of decoupled chains ($r=1$) as a product state of MPS with bond dimension $D_x$ for each individual spin-1 chain. In Fig.~\ref{fig:dx1} we show the  breaking of expected SU(2) symmetry through non-vanishing expectation values of $m$, $II_Q$, and $III_Q$ as function of $1/D_x$. This computation reveals that $D_x\ge9$ is necessary to obtain a good approximation of the ground state even in this simple limit, since  iPEPS with smaller $D_x$ display substantial breaking of SU(2) either by developing large magnetization or higher moments $II_Q$, $III_Q$.       
% \begin{figure}[tb]
%     \centering
%     \includegraphics[width=\columnwidth]{d3-d19_chi1_theta-neg1.96.eps}
%     \includegraphics[width=\columnwidth]{d3-d19_chi1_theta-neg1.18.eps}
%     \caption{\footnotesize{Spurious SU(2)-symmetry breaking induced by  finite-$D$ effects in dimerized phase of decoupled chains, $r=0$, for $\theta=-5/8\pi$ (top) and $\theta=-3/8\pi$ (bottom). Keeping $D_y=1$, for $D_x\ge9$ all order parameters $m$, $II_Q$, and $III_Q$ become small.}}
%     \label{fig:dx1}
% \end{figure}

\section{\label{appendix:methods_dmrg}Density matrix renormalization group (DMRG)}

\subsection{Subtleties in defining the N\'{e}el order parameters on finite systems}

In the main text, the order parameter for N\'{e}el phase is defined as the staggered magnetization, which is not an SU(2)-invariant observable. This will cause trouble when studying the SU(2) symmetry broken states in finite size systems because the SU(2) symmetry can only be broken in the thermodynamic limits. Specifically speaking, the DMRG algorithm that deals with the finite size cylinders does not have a preference on the choice of symmetry breaking "direction". As a result, in practice one may end up with states which are some linear superposition of different symmetry broken states, or even SU(2)-symmetric states. If the state is SU(2)-symmetric, the U(1)-symmetric staggered magnetization will apparently be vanishing. In our case, this situation corresponds to the simulations for the cut B ($\theta=-0.4375\pi$) which was initialized by the DVBS states (the simulations initialized by the N\'{e}el AFM states do not suffer from the problem since the initial states have already broken the SU(2) spin-rotation symmetry).  

To resolve this issue with DMRG on finite cylinders, we apply the pinning fields of N\'{e}el type at boundaries~\cite{dmrg2d} \iffalse (which were eventually turned off) \fi to explicitly break the global spin-rotation symmetry without disrupting too much the bulk physics. The pinning field strength is $|h_{\text{N\'{e}el}}^{\text{pin}}|=0.5$.

\subsection{Finite size scaling}
In the main context, we show the finite size scaling plot for cut B initialized by the N\'{e}el AFM state, in Fig.~\ref{fig:finite_size_scaling}. Here, we also attach the result for cut A with the FQ initial state. Given that the exponential decaying edge effects, we use an exponential fitting form versus the cylinder length $N_x$ introduced by~\cite{buch05} for finite size extrapolation in our analysis.
% \begin{figure}[tbh]
%     \flushleft
%     \includegraphics[scale=0.33]{plots/scaling_Nx_dimer_fq_dimerinit.png}
%     \caption{Finite size scaling of order parameters $D_h$ and $II_Q$ versus $N_x$ for cut A initialized by $|DVBS\rangle$.}
%     \label{fig:finite_size_scaling_dimer_fq_cut_dimerinit}
% \end{figure}
% \begin{figure}[H]
%     \flushleft
%     \includegraphics[scale=0.33]{plots/scaling_Nx_dimer_afm_dpninit.png}
%     \caption{Finite size scaling of order parameters $D_h$ and $m_s$ versus $N_x$ for cut A initialized by $|DVBS\rangle$.}
%     \label{fig:finite_size_scaling_dimer_afm_cut_dimerinit}
% \end{figure}
\begin{figure}[tbh]
    \flushleft
    \quad\includegraphics[scale=0.4]{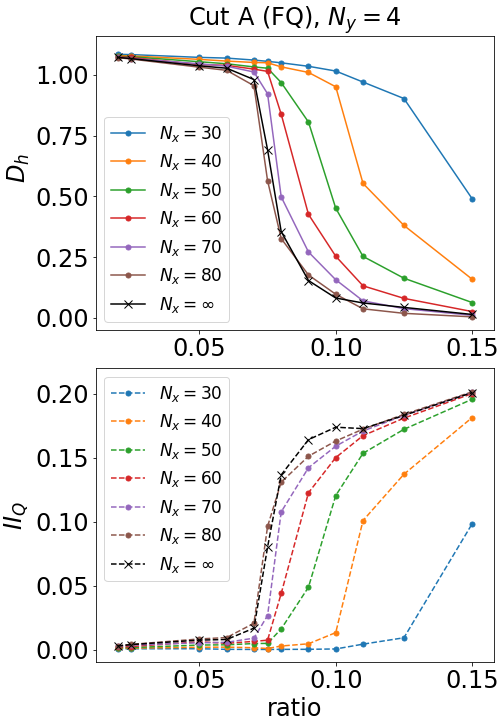}
    \caption{Finite size scaling of order parameters $D_h$ and $II_Q$ versus $N_x$ for cut A initialized by $|FQ\rangle$.}
    \label{fig:finite_size_scaling_dimer_fq_cut_fqinit}
\end{figure}

\begin{figure}[tb]
    \flushleft
    \quad\includegraphics[width=0.4\textwidth]{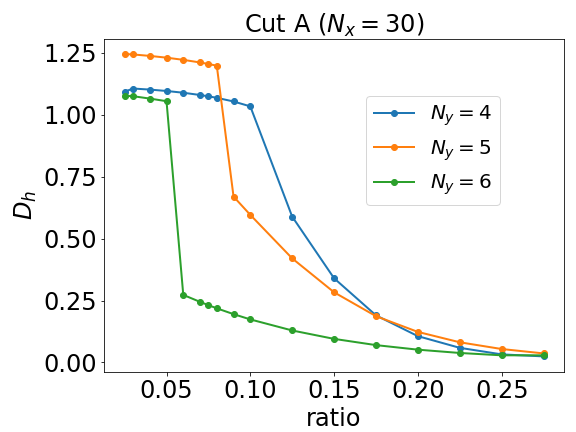}
    \caption{Melting of the dimerized order for different cylinder circumferences for cut A and fixed $N_x=30$.}
    \label{fig:finite_size_scaling_vs_Ny_dimer_fq_cut}
\end{figure}

In this work, DMRG simulations are performed systematically mainly for $N_y=4$, because we are more interested in the coupled chains behavior, and the computational cost of finite size scaling versus $N_y$ is very expensive. Here, we present how the dimerized order melts as ratio increases for fixed $N_x=30$ and different $N_y$'s (with $M=2000$ states kept, truncation errors $N_y=4,5,6$ simulations are $\varepsilon_{N_y=4}<2\times10^{-5}$, $\varepsilon_{N_y=5}<5\times10^{-5}$, $\varepsilon_{N_y=6}<3\times10^{-5}$), as shown by Fig.~\ref{fig:finite_size_scaling_vs_Ny_dimer_fq_cut}. As $N_y$ increases, the DVBS melting transition may happen at different ratios ($r=0.06~0.1$) for different $N_y$'s, but this do not qualitatively change our conclusion that the DVBS order melts very quickly.

\section{\label{sec:appendix_e_cross}Energy level crossings}
\begin{figure}[tbh]
    \flushleft
    \includegraphics[width=0.44\textwidth]{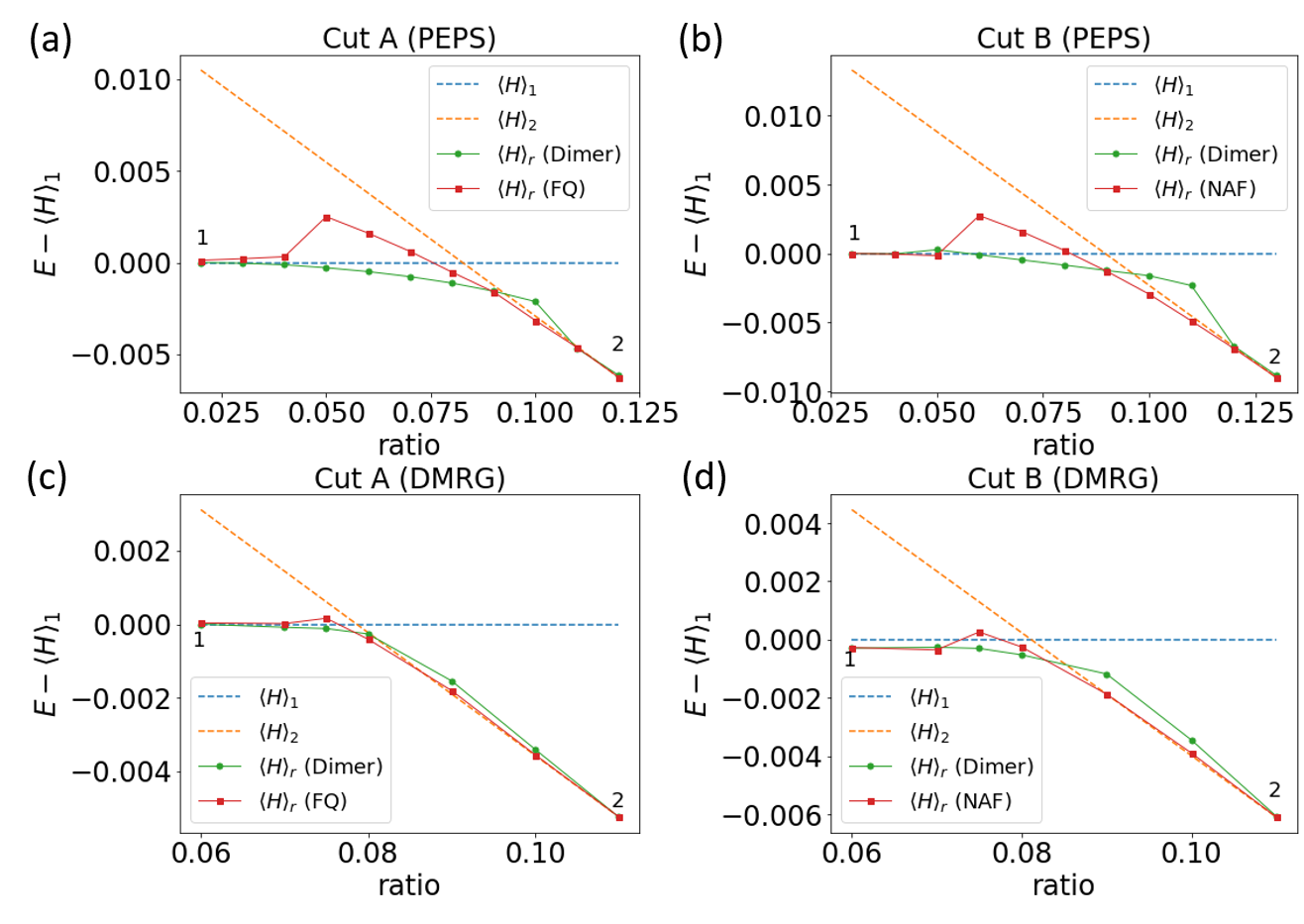}
    \caption{(a)-(d) Energy level crossings using different initial states for cuts A and B. $\langle H\rangle_{1,2}$ are computed using the state at $r_1$ initialized by $|DVBS\rangle$ at and the state at $r_2$ initialized by $|FQ\rangle$ for cut A or $|NAF\rangle$ for cut B.}
    \label{fig:e_level_crossing}
\end{figure}

In iPEPS and DMRG simulations, the simulations for each $\theta$ were initialized by two different states which represent different orders of the 1D- and 2D-limit phases. Apart from the hysterestic melting process of the DVBS order, we also observe the energy level crossings. By making use of this, we determine the phase boundary of the DVBS phase by the crossing points. Here, we show the energy level crossings (the energy difference) for both cuts A and B where both iPEPS and DMRG results are present, as shown by Fig.~\ref{fig:e_level_crossing}. Also, note that for DMRG results, we first extrapolated the energies in the $N_x\rightarrow\infty$ using the linear fitting function. The blue and orange dashed lines represent the "virtual" energies that are obtained by evaluating the Hamiltonian at some ratio $r$ yet using the states at $r_1$ and $r_2$. 

\section{\label{sec:appendix_corr_func} Correlation functions and correlation lengths}
We computed different types of correlation functions including the spin-spin, dimer-dimer and quadrupole-quadrupole  correlation functions ($C_S(r)$, $C_D(r)$ and $C_Q(r)$ ) for cuts A and B.

The spin-spin correlation is simply defined as
\vspace{-2mm}\begin{equation}
    C_{S}(r) = \left<\boldsymbol{S}(r)\cdot \boldsymbol{S}(0)\right>, 
\end{equation}

The dimer-dimer correlation function is defined as the product of two dimers, $D(0)$ and $D(r)$, given by  (note that they are different from the dimer order parameter $D_h$). The two dimers should not overlap with each other, which means that the position of the nearest dimer should be shifted by 2 with respect to the reference dimer.
\begin{equation}
    C_{D}(r)=\left<D((r+1)\boldsymbol{x})D(0)\right>,
\end{equation}
where the dimer is defined as
$D(r)=\boldsymbol{S}(\boldsymbol{r})\cdot\boldsymbol{S}(\boldsymbol{r}+\boldsymbol{x})$.

\begin{figure}[tbh]
    \centering
    \includegraphics[width=0.45\textwidth]{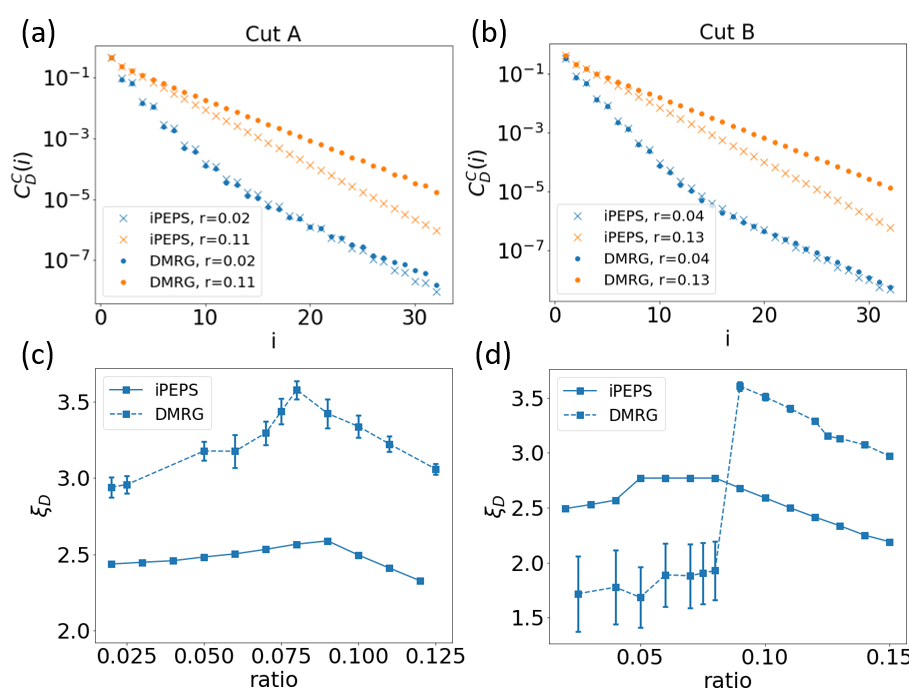}
    \caption{\footnotesize{(a,b) Connected dimer-dimer correlation functions at two ratios deeply in the two phases for cuts A and B. (c,d) Dimer-dimer correlation lengths for the two cuts.}}
    \label{fig:dd_corr}
\end{figure}

\begin{figure}[tbh]
    \centering
    \includegraphics[width=0.45\textwidth]{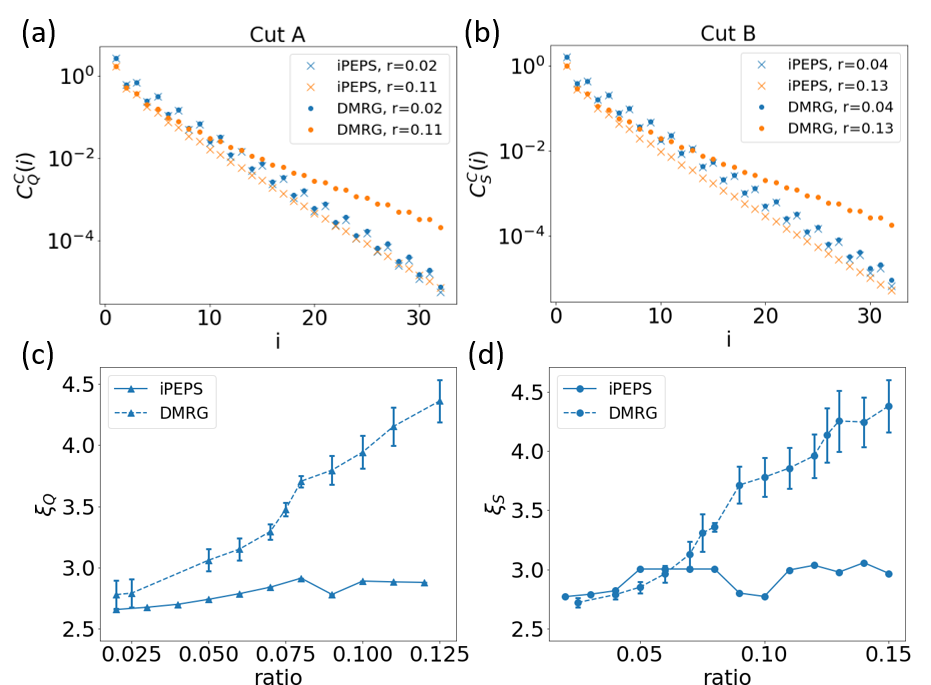}
    \caption{\footnotesize{(a,b) Connected quadrupole-quadrupole or spin-spin correlation functions at two ratios for the two cuts. (c,d) Quadrupole-Quadrupole and spin-spin correlation lengths for the two cuts.}}
    \label{fig:ssqq_corr}
\end{figure}

The quadrupole-quadrupole correlation function is given by 
\vspace{-2mm}\begin{equation}
    C_{Q}(r) = \left<\boldsymbol{Q}(r)\cdot \boldsymbol{Q}(0)\right>, 
\end{equation}
where the quadrupolar operator $\boldsymbol{Q}$, as a generalization of the spin vector for magnetic order, describes the quadrupolar order of spin-1 states, and is given by~\cite{Penc2011}
\begin{equation}
    \boldsymbol{Q} = \begin{pmatrix} 
    Q^{x^2-y^2} \\
    Q^{3z^2-r^2} \\
    Q^{xy} \\
    Q^{yz} \\
    Q^{zx} 
    \end{pmatrix} = \begin{pmatrix} 
    (S^x)^2-(S^y)^2 \\
    \frac{1}{\sqrt{3}}[3(S^z)^2 - S(S+1)] \\
    S^xS^y+S^yS^x \\
    S^yS^z+S^zS^y \\
    S^xS^z+S^zS^x 
    \end{pmatrix}
\end{equation}
To extract the correlation lengths, we compute the connected version of correlation functions, 
\begin{equation}
C^C_{\mathcal{O}}(r)=\left< \mathcal{O}(0)\cdot\mathcal{O}(r)\right> - \left< \mathcal{O}(0)\right>\cdot\left<\mathcal{O}(r)\right>
\end{equation}
(for the operator $\mathcal{O}=\boldsymbol{S}, \boldsymbol{Q},$ or $D$). We present the results of the connected correlation functions at two ratios for both cut A and cut B, as shown by Fig.~\ref{fig:dd_corr}(a,c) and Fig.~\ref{fig:ssqq_corr}(a,c). Then, the Ornstein-Zernike formula is used to extract the correlation lengths for both cuts, as shown in Fig.~\ref{fig:dd_corr}(b,d) and Fig.~\ref{fig:ssqq_corr}(b,d). In the vicinity of the phase boundary, we do not observe any singular behaviors for all the connected correlation function, excluding the possibility of continuous phase transition. The spin-spin and quadrupolar-quadrupolar correlation lengths are correspondingly finite in the DVBS phase, as shown in Fig.~\ref{fig:ssqq_corr}(c,d). The same is true of the dimer correlation length $\xi_D$ in Fig.~\ref{fig:dd_corr}(c,d), which is extracted from the long-distance tail of the dimer-dimer correlation function (after the short-range transient component decays). 

%It is worth mentioning that the piece-wise linear behavior seen in Fig.~\ref{fig:dd_corr}(a,c) for $\log{C^{C}_{D}(r)}$ can be explained by the presence of a sub-leading eigenvalue of the transfer matrix, $\lambda_n < \lambda_1$, such that the correlation function $C_D(r)=A_1 \lambda_1^r+A_n \lambda_n^r$ is dominated by the leading eigenvalue $\lambda_1$ at long distances, however with a crossover at short distances to a faster decay dictated by $\lambda_n$, provided a suitable coefficient $A_n>A_1$. Nevertheless, the physical correlation length, extracted in the long-wave-length limit is always given by the leading eigenvalue of the transfer matrix: 
%$\xi=-2[\ln{(\lambda_1)}]^{-1}$ (the extra factor $2$ originates from the choice of our unit cell). And we find $\lambda_1=0.48382$, $\xi=2.7545$ for $r=0.02$ on cut A as shown in Fig.~\ref{fig:dd_corr}(a), which is close to the value extracted from the connected correlation function, $\xi_{D,A}(r=0.02)=2.4928$.

\section{Mean-field result}
Consider the following Hamiltonian for general spin-S.
\begin{align}
    H =  \frac{1}{S^2}&\sum_{i,j} \bigg[J_\parallel\boldsymbol{S}_{i,j} \cdot \boldsymbol{S}_{i+1,j} + K_\parallel (\boldsymbol{S}_{i,j} \cdot \boldsymbol{S}_{i+1,j})^2 \notag\\ & +
    J_\perp\boldsymbol{S}_{i, j} \cdot \boldsymbol{S}_{i, j+1} + K_\perp (\boldsymbol{S}_{i, j} \cdot \boldsymbol{S}_{i,j+1})^2\bigg]
\end{align} 

where $J_\parallel=\cos{\theta}$, $K_\parallel=\sin{\theta}$, $J_\perp=r\cos{\theta}$, $K_\perp=r\sin{\theta}$. The biquadratic term can be expanded as follow.
\begin{align}
    (\boldsymbol{S}_1\cdot\boldsymbol{S}_2)^2 &=\uwave{\frac{1}{4}S_{1+}S_{1+}S_{2-}S_{2-}} +\underline{\frac{1}{4}S_{1+}S_{1-}S_{2-}S_{2+}}\notag\\
    &+\frac{1}{2}S_{1+}S_{1z}S_{2-}S_{2z}+\underline{\frac{1}{4}S_{1-}S_{1+}S_{2+}S_{2-}}\notag\\
    &+\uwave{\frac{1}{4}S_{1-}S_{1-}S_{2+}S_{2+}}+\frac{1}{2}S_{1-}S_{1z}S_{2+}S_{2z}\notag\\
    &+\frac{1}{2}S_{1z}S_{1+}S_{2z}S_{2-}+\frac{1}{2}S_{1z}S_{1-}S_{2z}S_{2+}\notag\\
    &+\underline{S_{1z}S_{1z}S_{2z}S_{2z}}
\end{align}

For N\'{e}el and dimerized state, only the underlined terms can have non-trivial contributions (\ref{eqn::mf_expr_naf},\ref{eqn::mf_expr_sn_ss_square},\ref{eqn::mf_expr_singlet_separate_ss_square}) because they do not change the onsite $S_z$ quantum numbers. For the $S=1$ SN state (FQ), if one chooses $|x\rangle$ or $|y\rangle$, the wavy underlined terms will also contribute (\ref{eqn::mf_expr_sn_ss_square_spin1}).\\

\noindent
\textbf{N\'{e}el AFM}
\begin{align}
    \langle S\overline{S}|\boldsymbol{S}_1\cdot \boldsymbol{S}_2 |S\overline{S}\rangle &= -S^2\notag\\
    \langle S\overline{S}|(\boldsymbol{S}_1\cdot \boldsymbol{S}_2)^2 |S\overline{S}\rangle &= \frac{1}{4}\sqrt{2S}\sqrt{2S}+S^2(-S)^2\notag\\
    =S^4+S^2
    \label{eqn::mf_expr_naf}
\end{align}

% FQ
% \begin{align}
%     \langle 00|\boldsymbol{S}_1\cdot \boldsymbol{S}_2 |00\rangle &= 0\notag\\
%     \langle 00|(\boldsymbol{S}_1\cdot \boldsymbol{S}_2)^2 |00\rangle &= 2\times\frac{1}{4}\sqrt{S(S+1)}\sqrt{S(S+1)}\notag\\
%     =\frac{1}{2}S^2(S+1)^2
%     \label{eqn::mf_expr_sn}
% \end{align}

\noindent
\textbf{Spin-nematic (FQ)} ($S\geq1$)
\begin{align}
    |\psi_{FQ}\rangle = |\phi^S\rangle = \frac{1}{\sqrt{2}}\big(|S\rangle+|\overline{S}\rangle\big)
    \label{eqn::sn_ansatz}
\end{align}
\begin{align}
    \langle \phi^S_{1} \phi^S_2|\boldsymbol{S}_1\cdot \boldsymbol{S}_2 |\phi^S_{1} \phi^S_2\rangle = 0
    \label{eqn::mf_expr_sn_ss}
\end{align}

$S=1$:
\begin{align}
    \langle \phi^S_{1} \phi^S_2|(\boldsymbol{S}_1\cdot \boldsymbol{S}_2)^2 |\phi^S_{1} \phi^S_2\rangle  &= 2\times\frac{1}{4}(\frac{2S}{2})^2+2\times\frac{1}{4}(\frac{2S}{2})^2+S^4\notag\\
    & =S^4+S^2=2
    \label{eqn::mf_expr_sn_ss_square_spin1}
\end{align}
$S>1$:
\begin{align}
    \langle \phi^S_{1} \phi^S_2|(\boldsymbol{S}_1\cdot \boldsymbol{S}_2)^2 |\phi^S_{1} \phi^S_2\rangle &= 2\times\frac{1}{4}(\frac{2S}{2})^2+S^4\notag\\
    =S^4+\frac{1}{2}S^2
    \label{eqn::mf_expr_sn_ss_square}
\end{align}

\noindent
\textbf{Dimerized VBS}

Note that there are two degenerate ground states for the dimerized (DVBS) order on the spin chain due to the translational symmetry breaking. The energy per site for the dimerized state should be averaged.

For the two spins from one dimer,
\begin{align}
|S_{tot}=0\rangle = \frac{1}{\sqrt{2S+1}}\sum_{m=-S}^{S}(-1)^{S-m}|m\rangle_1|S-m\rangle_2
\end{align}
Since $2\boldsymbol{S}_1\cdot \boldsymbol{S}_2 = (\boldsymbol{S}_1+ \boldsymbol{S}_2)^2-\boldsymbol{S}_1^2 - \boldsymbol{S}_2^2$, we have
\begin{align}
    \langle S_{tot}=0|\boldsymbol{S}_1\cdot \boldsymbol{S}_2 | S_{tot}=0\rangle &= -S(S+1)\notag\\
    \langle S_{tot}=0|(\boldsymbol{S}_1\cdot \boldsymbol{S}_2)^2 |S_{tot}=0\rangle &= S^2(S+1)^2
    \label{eqn::mf_expr_singlet}
\end{align}

For the two spins from two different dimers,
\begin{align}
    \langle S_{tot}^{13}=0|\langle S_{tot}^{24}=0|\boldsymbol{S}_1\cdot \boldsymbol{S}_2 |S_{tot}^{13}=0\rangle| S_{tot}^{24}=0\rangle = 0
    \label{eqn::mf_expr_singlet_separate_ss}
\end{align}
\begin{align}
    &\langle S_{tot}^{13}=0|\langle S_{tot}^{24}=0|(\boldsymbol{S}_1\cdot \boldsymbol{S}_2)^2 |S_{tot}^{13}=0\rangle| S_{tot}^{24}=0\rangle\notag\\
    =& \frac{2}{4} \times\bigg\{\frac{1}{2S+1}\sum_{m=-S}^{S-1}\big[S(S+1)-m(m+1)\big]\notag\\
    &\quad \times \frac{1}{2S+1}\sum_{m=-S+1}^{S}[S(S+1)-m(m-1)]\bigg\}\notag\\
    & +\bigg[\frac{1}{2S+1}\sum_{m=-S}^{S}m^2\bigg]^2=\frac{1}{3}S^2(S+1)^2
    \label{eqn::mf_expr_singlet_separate_ss_square}
\end{align}

\begin{figure}[tbh]
    \flushleft
    \includegraphics[scale=0.33]{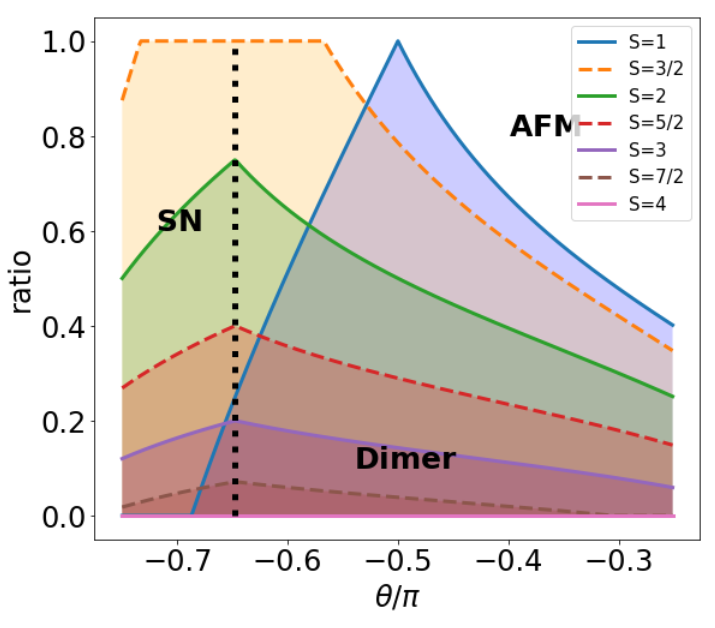}
    \caption{Mean-field phase diagram for spin-S BBH coupled chains with $\theta\in[-3\pi/4,-\pi/4]$. The shaded region indicates the dimerized phase. The solid lines indicate the boundaries for integer spins, and the dashed lines represent the boundaries for half integer spins.}
    \label{fig:mean_field_phase_higher_spins}
\end{figure}

With the information above, we then have the expressions for the mean-field ansatz. The mean-field phase diagram is shown by Fig.~\ref{fig:mean_field_phase_higher_spins}. 
\begin{align}
    E_{\text{NAF}}^{\text{mf}} &= -(1+r)\cos{\theta}+(S^2+1)(1+r)\sin{\theta}\\
    E_{\text{DVBS}}^{\text{mf}} &= -\frac{S+1}{2S}\cos{\theta}+\frac{(S+1)^2}{3}(2+r)\sin{\theta}\\
    E_{\text{FQ}}^{\text{mf}} &= \big(S^2+\frac{1}{2}+\frac{1}{2}\Theta(\frac{5}{4}-S)\big)(1+r)\sin{\theta}
\end{align}

\section{\label{appendix:methods_lfwt} Linear flavor wave theory}
Consider $S=1$ case. The Hamiltonian is given by
\begin{align}
    H =  &\sum_{i,j} \bigg[J_\parallel\boldsymbol{S}_{i,j} \cdot \boldsymbol{S}_{i+1,j} + K_\parallel (\boldsymbol{S}_{i,j} \cdot \boldsymbol{S}_{i+1,j})^2\bigg] \notag\\ & +
    \sum_{i, j} \bigg[J_\perp\boldsymbol{S}_{i, j} \cdot \boldsymbol{S}_{i, j+1} + K_\perp (\boldsymbol{S}_{i, j} \cdot \boldsymbol{S}_{i,j+1})^2\bigg]
\end{align} 

\subsection{Flavor wave boson representation (dipolar basis)}
In the dipolar ($S_z$) basis, the matrix form for operators $\boldsymbol{S}$ and $\boldsymbol{Q}$ are
\begin{align}
    &S^x=\frac{1}{\sqrt{2}}\begin{bmatrix}
    0 & 1 & 0\\
    1 & 0 & 1\\
    0 & 1 & 0
    \end{bmatrix},
    &S^y=\frac{1}{\sqrt{2}}\begin{bmatrix}
    0 & -i & 0\\
    i & 0 & -i\\
    0 & i & 0
    \end{bmatrix},\notag\\
    &S^z=\begin{bmatrix}
    1 & 0 & 0\\
    0 & 0 & 0\\
    0 & 0 & -1
    \end{bmatrix},
    &Q^{x^2-y^2}=\begin{bmatrix}
    0 & 0 & 1\\
    0 & 0 & 0\\
    1 & 0 & 0
    \end{bmatrix},\notag\\
    &Q^{3z^2-r^2}=\frac{1}{\sqrt{3}}\begin{bmatrix}
    1 & 0 & 0\\
    0 & -2 & 0\\
    0 & 0 & 1
    \end{bmatrix},    
    &Q^{xy}=\begin{bmatrix}
    0 & 0 & -i\\
    0 & 0 & 0\\
    i & 0 & 0
    \end{bmatrix},\notag\\
    &Q^{yz}=\frac{1}{\sqrt{2}}\begin{bmatrix}
    0 & -i & 0\\
    i & 0 & i\\
    0 & -i & 0
    \end{bmatrix},
    &Q^{zx}=\frac{1}{\sqrt{2}}\begin{bmatrix}
    0 & 1 & 0\\
    1 & 0 & -1\\
    0 & -1 & 0
    \end{bmatrix}
\end{align}

Switching to the bosonic language, the operators $\boldsymbol{S}$ and $\boldsymbol{Q}$ can be written in terms of the bosonic creation/annihilation operator, i.e.,
\begin{align}
    S_i^x &= \frac{1}{\sqrt{2}}(a_i^{1\dagger}a_i^{0}+a_i^{0\dagger}a_i^{1}+a_i^{0\dagger}a_i^{-1}+a_i^{-1\dagger}a_i^{0})\notag\\
    S_i^y &= \frac{1}{\sqrt{2}}(-ia_i^{1\dagger}a_i^{0}+ia_i^{0\dagger}a_i^{1}-ia_i^{0\dagger}a_i^{-1}+ia_i^{-1\dagger}a_i^{0})\notag\\
    S_i^z &= a_i^{1\dagger}a_i^{1} - a_i^{-1\dagger}a_i^{-1}\notag\\
    Q_i^{x^2-y^2} &= a_i^{1\dagger}a_i^{1} + a_i^{-1\dagger}a_i^{-1}\notag\\
    Q_i^{3z^2-r^2} &=  \frac{1}{\sqrt{3}}( a_i^{1\dagger}a_i^{1} + a_i^{-1\dagger}a_i^{-1} - 2a_i^{0\dagger}a_i^{0})\notag\\
    Q_i^{xy} &= -ia_i^{1\dagger}a_i^{-1} + ia_i^{-1\dagger}a_i^{1}\notag\\
    Q_i^{yz} &= \frac{1}{\sqrt{2}}(-ia_i^{1\dagger}a_i^{0}+ia_i^{0\dagger}a_i^{1}+ia_i^{0\dagger}a_i^{-1}-ia_i^{-1\dagger}a_i^{0})\notag\\
    Q_i^{zx} &= \frac{1}{\sqrt{2}}(a_i^{1\dagger}a_i^{0}+a_i^{0\dagger}a_i^{1}-a_i^{0\dagger}a_i^{-1}-a_i^{-1\dagger}a_i^{0})
\end{align}

\subsection{Flavor wave boson representation (quadrupolar basis)}
In the quadrupolar basis ($|\alpha\rangle,\alpha=x,y,z$), the matrix form for operators $\boldsymbol{S}$ and $\boldsymbol{Q}$ are
\begin{align}
    &S^x=\frac{1}{\sqrt{2}}\begin{bmatrix}
    0 & 0 & 0\\
    0 & 0 & -i\\
    0 & i & 0
    \end{bmatrix},
    &S^y=\frac{1}{\sqrt{2}}\begin{bmatrix}
    0 & 0 & i\\
    0 & 0 & 0\\
    -i & 0 & 0
    \end{bmatrix},\notag\\
    &S^z=\begin{bmatrix}
    0 & -i & 0\\
    i & 0 & 0\\
    0 & 0 & 0
    \end{bmatrix},
    &Q^{x^2-y^2}=\begin{bmatrix}
    -1 & 0 & 0\\
    0 & 1 & 0\\
    0 & 0 & 0
    \end{bmatrix},\notag\\
    &Q^{3z^2-r^2}=\frac{1}{\sqrt{3}}\begin{bmatrix}
    1 & 0 & 0\\
    0 & 1 & 0\\
    0 & 0 & -2
    \end{bmatrix},
    &Q^{xy}=\begin{bmatrix}
    0 & -1 & 0\\
    -1 & 0 & 0\\
    0 & 0 & 0
    \end{bmatrix},\notag\\
    &Q^{yz}=\frac{1}{\sqrt{2}}\begin{bmatrix}
    0 & 0 & 0\\
    0 & 0 & -1\\
    0 & -1 & 0
    \end{bmatrix},
    &Q^{zx}=\frac{1}{\sqrt{2}}\begin{bmatrix}
    0 & 0 & -1\\
    0 & 0 & 0\\
    -1 & 0 & 0
    \end{bmatrix}
\end{align}

Similar to the case in the dipolar basis, in the quadrupolar basis, the operators $\boldsymbol{S}$ and $\boldsymbol{Q}$ can be expressed as follows.
\begin{align}
    S_i^x &= -ia_i^{y\dagger}a_i^{z}+ia_i^{z\dagger}a_i^{y}\notag\\
    S_i^y &= -ia_i^{z\dagger}a_i^{x}+ia_i^{x\dagger}a_i^{z} \notag\\
    S_i^z &= -ia_i^{x\dagger}a_i^{y}+ia_i^{y\dagger}a_i^{x} \notag\\
    Q_i^{x^2-y^2} &= -a_i^{x\dagger}a_i^{x}+a_i^{y\dagger}a_i^{y}\notag\\
    Q_i^{3z^2-r^2} &=  \frac{1}{\sqrt{3}}( a_i^{x\dagger}a_i^{x} + a_i^{y\dagger}a_i^{y} - 2a_i^{z\dagger}a_i^{z})\notag\\
    Q_i^{xy} &= -a_i^{x\dagger}a_i^{y}-a_i^{y\dagger}a_i^{x}\notag\\
    Q_i^{yz} &= -a_i^{y\dagger}a_i^{z}-a_i^{z\dagger}a_i^{y}\notag\\
    Q_i^{zx} &= -a_i^{z\dagger}a_i^{x}-a_i^{x\dagger}a_i^{z}
\end{align}

\subsection{N\'{e}el phase}

For the N\'{e}el AFM state, we choose $|1\rangle$ for A sublattice and  $|-1\rangle$ for B sublattice. Then, we can replace $a_i^1$ and $b_i^{-1}$ by
\begin{align}
    a_i^{1}=a_i^{1\dagger} &=\sqrt{M-a_i^{0\dagger}a_i^{0}-a_i^{-1\dagger}a_i^{-1}}\notag\\
    &\approx\sqrt{M}-\frac{1}{2\sqrt{M}}a_i^{0\dagger}a_i^{0}-\frac{1}{2\sqrt{M}}a_i^{-1\dagger}a_i^{-1}\\
    b_i^{-1}=b_i^{-1\dagger} &=\sqrt{M-b_i^{0\dagger}b_i^{0}-b_i^{1\dagger}b_i^{1}}\notag\\
    &\approx\sqrt{M}-\frac{1}{2\sqrt{M}}a_i^{0\dagger}a_i^{0}-\frac{1}{2\sqrt{M}}a_i^{-1\dagger}a_i^{-1}
\end{align}

here $M=1$ for $S=1$. Then, we only keep terms up to the bilinear order for the interaction terms $\boldsymbol{S}_i\cdot\boldsymbol{S}_j$ and $\boldsymbol{Q}_i\cdot\boldsymbol{Q}_j=2(\boldsymbol{S}_i\cdot\boldsymbol{S}_j)^2+\boldsymbol{S}_i\cdot\boldsymbol{S}_j-8/3$. By doing such, we can get 

\begin{align}
    H_{\langle ij\rangle_\eta}^{AB} &=(J_\eta-K_\eta)(a_i^{0\dagger}a_i^{0}+a_i^{0\dagger}a_i^{0}+b_j^{0\dagger}b_j^{0}+a_i^{0}b_j^{0}+a_i^{0\dagger}b_j^{0\dagger})\notag\\
    &+(2J_\eta-K_\eta)(a_i^{-1\dagger}a_i^{-1}+b_j^{1\dagger}b_j^{1})\notag\\&+K_\eta(a_i^{-1}b_j^{1}+a_i^{-1\dagger}b_j^{1\dagger})+(-J_\eta+2K_\eta)
\end{align}

where $\eta=\parallel,\perp$. After that, we perform Fourier transformation $f_i=N^{-1}\sum_{\boldsymbol{k}}f_{\boldsymbol{k}}\exp(-i\boldsymbol{k}\cdot\boldsymbol{r}_i)$ and then obtain the energy with quantum fluctuation.
\begin{align}
    E=&E_{MF}^{0}+\sum_{\boldsymbol{k}}[a_{\boldsymbol{k}}^{0\dagger}, b_{-\boldsymbol{k}}^{0}]H_1(\boldsymbol{k})[a_{\boldsymbol{k}}^{0},b_{-\boldsymbol{k}}^{0\dagger}]\notag\\ 
    &+ [a_{\boldsymbol{k}}^{-1\dagger},b_{-\boldsymbol{k}}^{1}]H_2(\boldsymbol{k})[a_{\boldsymbol{k}}^{-1},b_{-\boldsymbol{k}}^{1\dagger}] -f_{1\boldsymbol{k}} - f_{2\boldsymbol{k}}
\end{align}
where 
\begin{align}
    H_i(\boldsymbol{k})=\begin{bmatrix}
    f_{i\boldsymbol{k}} & g_{i\boldsymbol{k}}\\
    g_{i\boldsymbol{k}} & f_{i\boldsymbol{k}}
    \end{bmatrix}, i=1,2
\end{align}
with 
\begin{align}
    f_{1\boldsymbol{k}}&=2(J_{\parallel}-K_{\parallel})+2(J_{\perp}-K_{\perp})\notag\\
    g_{1\boldsymbol{k}}&=2(J_{\parallel}-K_{\parallel})\cos{k_x}+2(J_{\perp}-K_{\perp})\cos{k_y}\notag\\
    f_{2\boldsymbol{k}}&=2(2J_{\parallel}-K_{\parallel})+2(2J_{\perp}-K_{\perp})\notag\\
    g_{2\boldsymbol{k}}&=2K_{\parallel}\cos{k_x}+2K_{\perp}\cos{k_y}
\end{align}

By doing the Bogoliubov diagonalization, the energy expression becomes
\begin{align}
    E=E_{MF}^{0}+\frac{1}{2}\sum_{\boldsymbol{k}}\big[&\omega_{1\boldsymbol{k}}(\alpha_{\boldsymbol{k}}^\dagger\alpha_{\boldsymbol{k}}+\beta_{\boldsymbol{k}}^\dagger\beta_{\boldsymbol{k}}+1)-f_{1\boldsymbol{k}}\notag\\+&\omega_{2\boldsymbol{k}}(\gamma_{\boldsymbol{k}}^\dagger\gamma_{\boldsymbol{k}}+\eta_{\boldsymbol{k}}^\dagger\eta_{\boldsymbol{k}}+1)-f_{2\boldsymbol{k}}\big]
\end{align}
where $\omega_{i\boldsymbol{k}}=\sqrt{f_{i\boldsymbol{k}}^2-g_{i\boldsymbol{k}}^2}$ are the energy eigenvalues.

Thus, the correction to the mean-field energy is
\begin{align}
    \Delta E&=\frac{1}{2}\sum_{\boldsymbol{k}}{(\omega_{1\boldsymbol{k}}+\omega_{2\boldsymbol{k}}-f_{1\boldsymbol{k}}-f_{2\boldsymbol{k}})}\notag\\&=\frac{1}{2}\frac{1}{(2\pi)^2}\int_{BZ}d\boldsymbol{k}(\omega_{1\boldsymbol{k}}+\omega_{2\boldsymbol{k}}-f_{1\boldsymbol{k}}-f_{2\boldsymbol{k}})
\end{align}

\subsection{FQ phase}
For the FQ case, we choose $|z\rangle$ for all the sites. Then, we can replace $a_i^z$ by
\begin{align}
    a_i^{z}=a_i^{z\dagger}&=\sqrt{M-a_i^{x\dagger}a_i^{x}-a_i^{y\dagger}a_i^{y}}\notag\\&\approx\sqrt{M}-\frac{1}{2\sqrt{M}}a_i^{x\dagger}a_i^{x}-\frac{1}{2\sqrt{M}}a_i^{y\dagger}a_i^{y}
\end{align}

here $M=1$ for $S=1$. Then, again, we only keep terms up to the bilinear order, O(M), for the interaction terms and perform the Fourier transform. In this way, one can obtain the energy expression in momentum space.
\begin{align}
    E=E_{MF}^0+\frac{1}{2}\sum_{\boldsymbol{k}}\big[f_{\boldsymbol{k}}(a_{\boldsymbol{k}}^{x\dagger}a_{\boldsymbol{k}}^{x}+a_{-\boldsymbol{k}}^{x\dagger}a_{-\boldsymbol{k}}^{x}+a_{\boldsymbol{k}}^{y\dagger}a_{\boldsymbol{k}}^{y}+a_{-\boldsymbol{k}}^{y\dagger}a_{-\boldsymbol{k}}^{y})\notag\\+g_{\boldsymbol{k}}(a_{\boldsymbol{k}}^{x\dagger}a_{-\boldsymbol{k}}^{x\dagger}+a_{\boldsymbol{k}}^{x}a_{-\boldsymbol{k}}^{x}+a_{\boldsymbol{k}}^{y\dagger}a_{-\boldsymbol{k}}^{y\dagger}+a_{\boldsymbol{k}}^{y}a_{-\boldsymbol{k}}^{y})\big]
\end{align}
with
\begin{align}
    f_{\boldsymbol{k}}&=-2(K_{\parallel}+K_{\perp})+2(J_{\parallel}\cos{k_x}+K_{\perp}\cos{k_y})\notag\\
    g_{\boldsymbol{k}}&=2(-J_{\parallel}+K_{\parallel})\cos{k_x}+2(-J_{\perp}+K_{\perp})\cos{k_y}
\end{align}
By doing the Bogoliubov diagonalization, we get
\begin{align}
    E=E_{MF}^{0}+\frac{1}{2}\sum_{\boldsymbol{k}}\big[\omega_{\boldsymbol{k}}(\alpha_{\boldsymbol{k}}^\dagger\alpha_{\boldsymbol{k}}+\alpha_{\boldsymbol{k}}^\dagger\alpha_{\boldsymbol{k}}+1)-f_{\boldsymbol{k}}\notag\\+\beta_{2\boldsymbol{k}}(\beta_{\boldsymbol{k}}^\dagger\gamma_{\boldsymbol{k}}+\beta_{\boldsymbol{k}}^\dagger\beta_{\boldsymbol{k}}+1)-f_{\boldsymbol{k}}\big]
\end{align}
where $\omega_{\boldsymbol{k}}=\sqrt{f_{\boldsymbol{k}}^2-g_{\boldsymbol{k}}^2}$ are the energy eigenvalues.

Thus, the correction to the FQ mean-field energy is
\begin{align}
    \Delta E=\sum_{\boldsymbol{k}}{(\omega_{\boldsymbol{k}}-f_{\boldsymbol{k}})}=\frac{1}{(2\pi)^2}\int_{BZ}d\boldsymbol{k}(\omega_{\boldsymbol{k}}-f_{\boldsymbol{k}})
\end{align}

\end{document}